\documentclass[12pt,halfline,a4paper]{ouparticle}
\usepackage{caption}
\usepackage{graphicx}
\usepackage{float} 
\usepackage{hyperref,xr} 
\usepackage{natbib} 
\usepackage{subfigure}
\usepackage{titling}
\usepackage{natbib}
\usepackage{color}
\usepackage{amsfonts,amsmath,amssymb}
\usepackage{theorem}
\newtheorem{proposition}{Proposition}

\newtheorem{condition}{Condition}

\newtheorem{theorem}{Theorem}
\newtheorem{remark}{Remark}

\makeatletter 
\newcommand{\fdsy@scale}{1.0} 
\newcommand\fdsy@mweight@normal{Book} 
\newcommand\fdsy@mweight@small{Book} 
\newcommand\fdsy@bweight@normal{Medium} 
\newcommand\fdsy@bweight@small{Medium} 
\renewcommand\appendix{\par
	\setcounter{section}{0}
	\setcounter{subsection}{0}
	\gdef\thesection{Appendix \Alph{section}}}
\usepackage{algorithm}
\usepackage{algpseudocode}
\usepackage{algcompatible}

\DeclareFontFamily{U}{FdSymbolC}{} 
\DeclareFontShape{U}{FdSymbolC}{m}{n}
{ <-7.1> s * [\fdsy@scale] FdSymbolC-\fdsy@mweight@small
	<7.1-> s * [\fdsy@scale] FdSymbolC-\fdsy@mweight@normal 
}{} 
\DeclareFontShape{U}{FdSymbolC}{b}{n}
{ <-7.1> s * [\fdsy@scale] FdSymbolC-\fdsy@bweight@small
	<7.1-> s * [\fdsy@scale] FdSymbolC-\fdsy@bweight@normal
}{} 
\DeclareSymbolFont{arrows}{U}{FdSymbolC}{m}{n}
\SetSymbolFont{arrows}{bold}{U}{FdSymbolC}{b}{n} 
\DeclareMathSymbol{\upvDash}{\mathrel}{arrows}{233} 
 
\DeclareMathSymbol{\upmodels}{\mathrel}{arrows}{237} 
\makeatother
\begin{document}
	\predate{}
	\postdate{}
	\title{Transfer Learning for High-dimensional Quantile Regression via Convolution Smoothing}
	
	\author{\name{Yijiao Zhang and Zhongyi Zhu}\address{\it Department of Statistics and Data Science, Fudan University}}

	\date{} 
	\maketitle
	\begin{quotation}
		\begin{center}
			{ \textbf{Abstract}}
		\end{center}
	{\bf }
		This paper studies the high-dimensional quantile regression problem under the transfer learning framework, where possibly related source datasets are available to make improvements on the estimation or prediction based solely on the target data. In the oracle case with known transferable sources, a smoothed two-step transfer learning algorithm based on convolution smoothing is proposed and the $\ell_1$/$\ell_2$ estimation error bounds of the corresponding estimator are also established. To avoid including non-informative sources, we propose to select the transferable sources adaptively and establish its selection consistency under regular conditions.  Monte Carlo simulations as well as an empirical analysis of gene expression data demonstrate the effectiveness of the proposed procedure.

		\vspace{9pt}
		\par
	\end{quotation}\par

\section{Introduction}
\label{sec:intro}

The increasing availability of datasets from multiple sources has provided us with unprecedented opportunities to get a better understanding of the data-limited target problem. For example, for the task of drug sensitivity prediction, the drug response data for the target type of cancer may be limited, but source data for another cancer type may be sufficient \citep{turki2017transfer}. However, there is no free lunch. Along with the satisfactory sample size of source studies comes the heterogeneity between the source and the target. Intuitively, the more related the source to the target, the more improvement may be made in learning about the target. This motivates transfer learning \citep{torrey2010transfer,pan2009survey,weiss2016survey,niu2020decade}, which attempts to improve a learner from one domain by transferring information from a related but different domain. A considerable amount of research has clearly shown the success of transfer learning in many real-world applications, including ride dispatching \citep{wang2018deep}, medical images analyses \citep{yu2022transfer}, and human activity recognition \citep{hirooka2022ensembled}, etc. 

This paper aims to investigate the effect of transfer learning on quantile regression (QR) in a high-dimensional setting. Ever since the influential work of \cite{koenker1978regression}, numerous scholars have explored the theoretical characterization of QR in various areas. See \cite{Koenker2017quantile} for a detailed review. Compared to the traditional conditional mean regression, QR captures the heterogeneous impact of regressors on different parts of the distribution. It also exhibits robustness to heteroscedastic and heavy-tailed errors. Moreover, in view of the frequently-collected high-dimensional data in various application domains including genomics, tomography, and finance, we focus on the regime where the dimension of covariates is substantially larger than the sample size. 

A limited number of studies have sought to examine the possibility of interaction between quantile regression and multiple sources of data. Among them, \cite{fan2016multiqr} considered the multi-task quantile regression problem under the transnormal model, which may be restricted in practice. Moreover, large-scale covariance matrix estimation is needed in their method. In the field of data integration, \cite{dai2023di} proposed to use multiple datasets together with multiple quantiles to select variables simultaneously. They established model selection consistency and asymptotic normality of their estimator, but theoretical results about the estimation error are left unknown. More importantly, it needs to be emphasized that, in contrast to multi-task learning and data integration, where the datasets are equally important, the roles of the source and target tasks are no longer symmetric in transfer learning since we care most about the performance on the target data.

Despite the popularity of transfer learning on the practical side, statistical views of it as well as the theoretical underpinnings of related algorithms have been less than satisfactory. In the context of nonparametric classification, \citet{cai2021transfer} proposed an adaptive classifier and established its minimax optimality. For high-dimensional data analysis, existing research has focused on penalized mean regression.  \cite{bastani2021predicting} considered linear regression and derived the estimation error bound for transferring knowledge from a single source to the target. \cite{li2022trans} investigated the case of multiple sources and proved the minimax optimal rate of their estimator. \citet{tian2022transfer} further extended their work to generalized linear models and provided analogous theoretical guarantees. The benefits for confidence interval construction are also studied. These studies used a two-step framework, which was further applied to Gaussian graphical models with false discovery rate control \citep{Li2022transggm,he2021transfer} and functional learning models \citep{lin2022transfer}. By measuring the similarity between the source and target by the difference of corresponding regression coefficients, \citet{li2022trans} and \citet{tian2022transfer} also allow some of the source studies to be non-informative. However, one restriction of these studies is that they assume homoscedastic random errors with sub-gaussian distributions. While in real life, it is common to observe heteroscedastic variance \citep{delaigle2007nonparametric,delaigle2008density} and heavy-tailed
noises \citep{fan2017estimation,sun2020adaptive}. Furthermore, the similarity between the source and target may vary across different tails of the outcome distribution, which can not be reflected by the mean regression model. 

Motivated by the above concerns, given a target dataset $\{(({\boldsymbol{x}}_{i}^{(0)})^{\top},y_{i}^{(0)})\}_{i=1}^{n_0}$ and $K$ independent source datasets $\{\{(({\boldsymbol{x}}_{i}^{(k)})^{\top},y_{i}^{(k)})\}_{i=1}^{n_k}\}_{k=1}^{K}$, we investigate the conditional quantile of the response $y_i^{(k)}$ conditional on
the covariates ${\boldsymbol{x}}_{i}^{(k)}\in\mathbb{R}^{p}$ at a given quantile level $\tau\in(0,1)$, denoted by $\mathcal{Q}_{y_{i}^{(k)}\mid {\boldsymbol{x}}_{i}^{(k)}}(\tau)$. We consider a linear QR model, that is, $\mathcal{Q}_{y_{i}^{(k)}\mid {\boldsymbol{x}}_{i}^{(k)}}(\tau)=({\boldsymbol{x}}_{i}^{(k)})^{\top}{\boldsymbol{w}}^{(k)}(\tau)$ with ${\boldsymbol{w}}^{(k)}(\tau) \in \mathbb{R}^{p}$ being the coefficient vector. We omit the dependence of ${\boldsymbol{w}}^{(k)}(\tau)$ on $\tau$ hereafter. The preceding model can be rewritten as
\begin{equation}\label{eq:model}
\boldsymbol{y}^{(k)}={\boldsymbol{X}}^{(k)}{{\boldsymbol{w}}^{(k)}}+{\boldsymbol{\epsilon}}^{(k)}, \quad k=0,\ldots, K,
\end{equation}	
where  
${\boldsymbol{y}}^{(k)}=(y^{(k)}_{1},\ldots,y^{(k)}_{n_k})^{\top}\in \mathbb{R}^{n_k}$, ${\boldsymbol{X}}^{(k)}=({\boldsymbol{x}}_{1}^{(k)}, \ldots, {\boldsymbol{x}}^{(k)}_{n_k})^{\top}\in\mathbb{R}^{n_k\times p}$ is the design matrix,  and ${\boldsymbol{\epsilon}}^{(k)}\in \mathbb{R}^{n_k}$ is the noise vector with the $i$-th element ${\epsilon}^{(k)}_{i}$ satisfying $\mathbb{P}({\epsilon}^{(k)}_{i}\leq 0\mid {\boldsymbol{x}}^{(k)}_{i})=\tau$. 
The target parameter $\boldsymbol{\beta}:=\boldsymbol{w}^{(0)}$, which satisfies $\|\beta\|_{0}=s\ll p$, is of our primary interest. Besides, we do not assume $\ell_0$-sparsity on $\boldsymbol{w}^{(k)}$ for $k=1,\ldots,K$. Following the spirit of \cite{li2022trans} and \cite{tian2022transfer}, we define the $k$-th contract vector ${\boldsymbol{\delta}}^{(k)}=\boldsymbol{\beta}-{\boldsymbol{w}}^{(k)}$ and the transferability of the $k$-th study as $\|{\boldsymbol{\delta}}^{(k)}\|_1$. In general, we prefer $\|{\boldsymbol{\delta}}^{(k)}\|_1$ to be sufficiently small to guarantee performance improvement of transfer learning. With this insight in mind, the transferable set of source studies is defined as $\mathcal{A}_{\eta}=\left\{1 \leq k \leq K:\left\|\boldsymbol{\delta}^{(k)}\right\|_{1} \leq \eta\right\}$. 

Compared with existing works for the (generalized) linear models under the transfer learning framework \citep{li2022trans,tian2022transfer}, our proposed QR model has several advantages:
\begin{enumerate}
	\item it offers a more complete picture of the target problem by varying $\tau$;
	\item it could handle heterogeneity due to either heteroscedastic variance or other forms of non-location-scale covariate effects;
	\item it relaxes the distributional conditions on the error terms, which is more robust to tail behavior;
	\item it allows the difference ${\boldsymbol{\delta}}^{(k)}$ as well as the transferable set $\mathcal{A}_{\eta}$ to change with $\tau$, which is more flexible.
\end{enumerate}

However, the adoption of QR plays the role of a double-edged sword. More specifically, under the transfer learning framework, three critical issues need to be addressed: (i) \textit{what to transfer}, (ii) \textit{how to transfer}, and (iii) \textit{when to transfer} \citep{pan2009survey}. The first one can be similarly solved by defining some common component as that in \citet{li2022trans} and \citet{tian2022transfer}. A natural solution to the second issue is to adopt the two-step framework. Unfortunately, due to the non-differentiability of the quantile loss function, this can be not only technically challenging but also computationally expensive, especially in the transfer learning setting when the auxiliary source data may be extremely large. Besides, the (possibly) unrelated tasks pose additional challenges to the last issue, which involves correctly identifying the transferable set $\mathcal{A}_{\eta}$ to circumvent negative transfer \citep{zhang2020survey,niu2020decade}.

To tackle the aforementioned challenges, we combine a smoothed two-step procedure with a source detection algorithm for transferring high-dimensional QR models. To deal with the non-smoothness, we employ a recently developed convolution smoothing technique  \citet{fernandes2021smoothing,he2021smoothed} to smooth the piecewise linear quantile loss. Convolution smoothing and convex relaxation enable us to use gradient-based algorithms which are much more scalable to large-scale datasets. At the same time, delicated analysis of the smoothing bandwidths is also needed to control the smoothing bias. A distributed QR transfer approach is also proposed for the ease of computation burden. To avoid negative transfer, we propose to evaluate the change of performance on the target with and without each source and exclude unrelated sources which lead to worse performance. This idea originates from the works by \citet{Eaton2008modeling} and \cite{tian2022transfer}. 

In theory, our contributions are fourfold. Firstly, by choosing smoothing bandwidths properly such that the smoothing bias is controlled, we derived the $\ell_{1}$ and $\ell_2$ estimation error bounds for our smoothed two-step estimator, which is complementary to the existing results in the mean regression world. The convergence rate is shown to be faster than that of the single-task smoothed high-dimensional quantile regression established in \cite{tan2022conv}. Secondly, we show that the transferable set can be consistently detected by the proposed clustering method when there is a sufficiently large gap between the positive and negative sources. Thirdly, the statistical property of our distributed QR transfer estimator is also established. As a byproduct, two lemmas related to the local restricted strong convexity (RSC) of the empirical smoothed quantile loss are established, which provide a core result for establishing error bounds for our smoothed two-step QR estimators and may be of independent interest.

The rest of the paper is organized as follows. In Section \ref{sec:method}, we present our smoothed two-step estimator with the known transferable set, followed by a source detection algorithm to select the transferable set. Section \ref{sec:theory} is dedicated to providing theoretical guarantees for our proposed method, including estimation error bounds and the selection consistency of the source detection algorithm. We demonstrate our proposed method on simulated data in Section \ref{sec:simu} and an empirical analysis of Genotype-Tissue Expression (GTEx) data in Section \ref{sec:appl}. We conclude with some discussions on possible extensions in Section \ref{sec:disc}. Additional simulation results and an extension to the distributed QR transfer are relegated to the Appendix.

We finish this section with notation. Throughout this paper, we use bold capitalized letters (e.g. $\boldsymbol{X}, \boldsymbol{A})$ to denote matrices and use bold little letters (e.g. $\boldsymbol{x}$, $\boldsymbol{y})$ to denote vectors. For a $p$-dimensional vector $\boldsymbol{x}=\left(x_{1}, \ldots, x_{p}\right)^{T}$, we denote its $\ell_{q}$-norm as $\|\boldsymbol{x}\|_q=\left(\sum_{i=1}^{p}\left|x_{i}\right|^{q}\right)^{1 / q}(q \in(0,2])$, $\ell_{0}$-norm as $\|\boldsymbol{x}\|_{0}=\#\left\{j: x_{j} \neq 0\right\}$ and $\ell_{\infty}$-norm as $\|\boldsymbol{x}\|_{\infty}$=$\max_{j}|x_j|$. We use $|\mathcal{I}|$ to denote the cardinal number of a set $\mathcal{I}$. For any positive integer, $n$, we denote the index set $\{1,\ldots,n\}$ as $[n]$. We use $\mathbf{I}\{\cdot\}$ to denote the indicator function. For a matrix $\boldsymbol{A}_{p \times q}=\left[a_{i j}\right]_{m \times n}$, let $\|\boldsymbol{A}\|_1=\sup _j \sum_{i=1}^m\left|a_{i j}\right|$ and 
$\|\boldsymbol{A}\|_{\max }=\sup _{i, j}\left|a_{i j}\right|$ denote its $\ell_1$-norm and max-norm respectively. For any $k \times k$ symmetric, positive semidefinite matrix $\mathbf{A} \in \mathbb{R}^{m \times m}$, we use $\Lambda_{\mathrm{min}}(\mathbf{A})$ and $\Lambda_{\mathrm{max}}(\mathbf{A})$ to denote its minimum and maximum eigenvalue. For any two real numbers $a$ and $b$, we write $a \vee b=\max (a, b)$ and $a \wedge b=\min (a, b) .$ For two sequences of non-negative numbers $\left\{a_n\right\}_{n \geq 1}$ and $\left\{b_n\right\}_{n \geq 1}$, $a_n \lesssim b_n$ or $b_n\gtrsim a_n$ indicate that $\sup _n\left|a_n / b_n\right|<\infty$; $a_n \asymp b_n$ is equivalent to $a_n \lesssim b_n$ and $b_n \lesssim a_n$. We use $a_n \ll b_n, b_n \gg a_n$ or $a_n=o\left(b_n\right)$ to represent $\left|a_n / b_n\right| \rightarrow 0$ as $n \rightarrow \infty$. 

\section{Methodology}
\label{sec:method}
\subsection{QR basics}
\label{subsec:QR}
We start by providing a brief introduction to QR. Consider the QR model
\begin{equation}\label{eq:qr}
{y_i}={{\boldsymbol{x}}_{i}}^{\top}{{\boldsymbol{w}}}+{\epsilon_i}, \quad \mathbb{P}({\epsilon}_{i}\leq 0\mid {\boldsymbol{x}}_{i})=\tau, \quad i=1,\ldots, n
\end{equation} 
where $\boldsymbol{w}\in\mathbb{R}^p$ is the coefficient vector with $\|\boldsymbol{w}\|_0=s$. Following \cite{belloni2011L1}, we can estimate $\boldsymbol{w}$ by fitting the $\ell_1$-penalized quantile regression ($\ell_1$-QR):
\begin{equation}\label{eq:l1qr}
\tilde{\boldsymbol{w}}\in\underset{\boldsymbol{w} \in \mathbb{R}^{p}}{\operatorname{argmin}}\left\{\frac{1}{n}\sum_{i=1}^{n}{\rho}_{\tau}\left({y}_i-{\boldsymbol{x}_i}^{\top}\boldsymbol{w}\right)+\lambda_{\boldsymbol {w}}\|\boldsymbol{w}\|_{1}\right\},
\end{equation} 
where $\rho_{\tau}(x) = x[\tau-\mathbf{I}\{x \leq 0\}]$ is the check function. However, if we fit $\ell_1$-QR based solely on the target data $(\boldsymbol{X}^{(0)},\boldsymbol{y}^{(0)})$ to obtain an estimator $\tilde{\boldsymbol{\beta}}_{\text{tar}}$, its performance can be limited by the sample size $n_0$ of the target data, which motivates us to incorporate other related sources to make some improvements. 

\subsection{Smoothed two-step QR transfer}
\label{subsec:kqr}
We start with the case when the transferable set $\mathcal{A}_{\eta}$ is known. As there is heterogeneity between the target and source data, it leads us to think about the first issue in transfer learning: \textit{What to transfer}. The key point is to find a common component between the source and the target. In view of this, we pool the target and the sources in the transferable set $\mathcal{A}_{\eta}$ together and consider the parameter ${\boldsymbol{w}}^{\mathcal{A}_{\eta}}$ identified by the following moment equation
\begin{equation}\label{eq:wAmoment}
\sum_{k \in \mathcal{A}_{\eta}\cup\{0\}}\alpha_{k}\mathbb{E}\left[\boldsymbol{x}^{(k)}\left(F_{\epsilon^{(k)}\mid \boldsymbol{x}^{(k)}}\left({(\boldsymbol{x}^{(k)})}^{\top}{(\boldsymbol{w}^{\mathcal{A}_{\eta}}-\boldsymbol{w}^{(k)})}\mid \boldsymbol{x}^{(k)}\right)-\tau\right)\right]=0,
\end{equation}
where $ \alpha_{k}={n_{k}}/{(n_{\mathcal{A}_{\eta}}+n_0)}$ with  $n_{\mathcal{A}_{\eta}}=\sum_{k\in\mathcal{A}_{\eta}}n_k$ and $F_{\epsilon^{(k)}\mid \boldsymbol{x}^{(k)}}(\cdot\mid \boldsymbol{x}^{(k)})$ is the conditional distribution function of $\epsilon^{(k)}$ given $\boldsymbol{x}^{(k)}$. Further denote $f_{\epsilon^{(k)}\mid \boldsymbol{x}^{(k)}}(\cdot)$ as the conditional density of $\epsilon^{(k)}$ given $\boldsymbol{x}^{(k)}$, and $\boldsymbol{M}_k=\mathbb{E}\left[\int_{0}^{1}\boldsymbol{x}^{(k)}{(\boldsymbol{x}^{(k)})}^{\top}f_{\epsilon^{(k)}\mid \boldsymbol{x}^{(k)}}(t{(\boldsymbol{x}^{(k)})}^{\top}(\boldsymbol{w}^{\mathcal{A}_{\eta}}-\boldsymbol{w}^{(k)}))\mathrm{dt}\right]$. Then we can write
\begin{equation*}
{\boldsymbol{w}^{\mathcal{A}_{\eta}}}={\boldsymbol\beta} -{\boldsymbol\delta^{\mathcal{A}_{\eta}}},
\end{equation*}
where $\boldsymbol\delta^{\mathcal{A}_{\eta}}=\left(\sum_{k \in \mathcal{A}_{\eta}\cup\{0\}} \alpha_{k} \boldsymbol{M}_k\right)^{-1} \sum_{k \in \mathcal{A}_{\eta}\cup\{0\}} \alpha_{k} \boldsymbol{M}_k \boldsymbol\delta^{(k)}$. The bias $\boldsymbol\delta^{\mathcal{A}_{\eta}}$ is a weighted average of $\boldsymbol\delta^{(k)}$, which is supposed to be $\ell_{1}$-sparse under certain conditions. Since the moment equation (\ref{eq:wAmoment}) incorporates information from the sources and the target, ${\boldsymbol{w}}^{\mathcal{A}_{\eta}}$ can be seen as a shared knowledge which can be transferred. 

Now we move to consider the second issue: \textit{how to transfer}. A commonly used heuristic in transfer learning is to adopt the two-step framework \citep{bastani2021predicting,li2022trans,tian2022transfer}. Specifically, in the first transferring step, we can pool the sources and target together to obtain an estimator $\hat{\boldsymbol{w}}^{\mathcal{A}_{\eta}}$ and then further correct the bias of $\hat{\boldsymbol{w}}^{\mathcal{A}_{\eta}}$ by calibrating it using the target data. Both steps could be accomplished by fitting $\ell_1$-QR similar to that in (\ref{eq:l1qr}). However, the nondifferentiable quantile loss function brings challenges to both computation and theory establishment.

On the computation side, the minimization problem in (\ref{eq:l1qr}) can be reformulated as a linear program \citep{wang2012quantile,peng2015iterative} of which the computation complexity grows with both $p$ and $n$. In the setting of multiple sources as well as high-dimensional data, this approach may suffer from heavy computational costs. On the theoretical side, due to the nonsmooth loss and (possibly) non-sparsity of $\hat{\boldsymbol{w}}^{\mathcal{A}_{\eta}}$, it is difficult to establish theoretical guarantees. 

To overcome these challenges, we propose a smoothed version of the two-step procedure by smoothing the piecewise linear quantile loss via convolution. Recall the QR model in (\ref{eq:qr}). The $\ell_1$-penalized smoothed quantile regression ($\ell_1$-SQR) estimator \citep{tan2022conv} is defined as
\begin{equation}\label{eq:l1sqr}
\hat{\boldsymbol{w}}\in\underset{\boldsymbol{w} \in \mathbb{R}^{p}}{\operatorname{argmin}}\left\{\frac{1}{nh}\sum_{i=1}^{n}\int_{-\infty}^{\infty}{\rho}_{\tau}(u)K\left(\frac{u+{(\boldsymbol{x}_{i})}^{\top}\boldsymbol{w}-{y}_{i}}{h}\right)\mathrm{d}u+\lambda\|\boldsymbol{w}\|_{1}\right\},
\end{equation} 
where $h$ is the smoothing bandwidth, $\lambda$ is a tuning parameter that controls the model complexity, and $K: \mathbb{R} \rightarrow[0, \infty)$ is a symmetric, non-negative kernel that integrates to one. For ease of notation, we use $\ell_{1}$-SQR($\{(\boldsymbol{x}_{i},y_i)\}_{i\in\mathcal{I}};\lambda,h$) to denote the estimator in (\ref{eq:l1sqr}) given a dataset $\{(\boldsymbol{x}_{i},y_i)\}_{i\in\mathcal{I}}$ with index set $\mathcal{I}$, a tuning parameter $\lambda$ and a bandwidth $h$.

The convolution-type smoothing first introduced by \citet{fernandes2021smoothing} yields a convex and twice differentiable loss, which enables the use of gradient-based algorithms and hence eases the computational burden. Additionally, it is also convenient for statistical analysis. Consider the single task quantile regression based solely on the target data $(\boldsymbol{X}^{(0)},\boldsymbol{y}^{(0)})$. In the low-dimensional regime where $p \ll n_0$, \citet{fernandes2021smoothing} provided a comprehensive asymptotic analysis for the unpenalized smoothed QR estimator, followed by an in-depth finite sample theory in \citet{he2021smoothed}. More related to our work, \citet{tan2022conv} investigated convolution smoothing for high-dimensional QR on both the theoretical and computational sides. They showed that with a proper yet flexible choice of the bandwidth, the $\ell_1$-SQR estimator shares the same $\ell_{1}$ and $\ell_{2}$ error upper bounds as the $\ell_1$-QR estimator \citep{belloni2011L1}, which are $s\sqrt{\log p/n_0}$ and $\sqrt{s\log p/n_0}$ respectively. In addition, a coordinate descent algorithm and an alternating direction method of multiplier algorithm were also proposed for solving the $\ell_1$-SQR problem in (\ref{eq:l1sqr}) with the uniform kernel and Gaussian kernel respectively.

Combining this smoothing procedure with our two-step framework leads to Algorithm \ref{alg:otqr} ( Oracle-Trans-SQR). Here $\mathcal{I}_k=[n_k]$ for $k=0,\ldots,K$. The smoothed two-step estimator $\hat{\boldsymbol{\beta}}$ is expected to converge fast to our target $\boldsymbol{\beta}$ provided a substantial sample size of source studies in the transferring step along with a sufficiently small bias in the debiasing step.

\begin{algorithm}[h]
	\caption{Oracle-Trans-SQR Algorithm}
	\label{alg:otqr} 
	\begin{algorithmic}[1]
		\REQUIRE Target data $(\boldsymbol{X}^{(0)}, \boldsymbol{y}^{(0)})$, source data $\{(\boldsymbol{X}^{(k)}, \boldsymbol{y}^{(k)})\}_{k \in \mathcal{A}_{\eta}}$, penalty parameters $(\lambda_{\boldsymbol{w}},\lambda_{\boldsymbol{\delta}})$ and bandwidths $(h_{\boldsymbol{w}},h_{\boldsymbol{\delta}})$.
		\STATE \underline{\textbf{Transferring :}} 
		$\hat{\boldsymbol{w}}^{\mathcal{A}_{\eta}}\leftarrow \ell_{1}\text{-SQR}(\{\{(\boldsymbol{x}^{(k)}_{i},y^{(k)}_i)\}_{i\in\mathcal{I}_k}\}_{k\in\mathcal{A}_{\eta}\cup\{0\}};\lambda_{\boldsymbol{w}},h_{\boldsymbol{w}})$
		
		\STATE \underline{\textbf{Debiasing :}} 
		${\hat{\boldsymbol{\delta}}^{\mathcal{A}_{\eta}}}\leftarrow \ell_{1}\text{-SQR}(\{(\boldsymbol{x}^{(0)}_{i},y^{(0)}_i-{(\boldsymbol{x}^{(0)}_{i})}^{\top}\hat{\boldsymbol{w}}^{\mathcal{A}_{\eta}})\}_{i\in\mathcal{I}_0};\lambda_{\boldsymbol{\delta}},h_{\boldsymbol{\delta}})$
		
		\ENSURE $ {\hat{\boldsymbol\beta}}={\hat{\boldsymbol{w}}^{\mathcal{A}_{\eta}}}+{\hat{\boldsymbol\delta}^{\mathcal{A}_{\eta}} }$
	\end{algorithmic}
\end{algorithm}

Due to the memory constraints of a single machine, we also provide a distributed version of Algorithm \ref{alg:otqr} with corresponding statistical properties, which are relegated to \ref{sec:distsqr}. 

\subsection{Transferable set detection}
\label{subsec:ukqr}
In practice, the transferable set $\mathcal{A}_{\eta}$ may be unknown, which makes the problem more intractable. Indeed, the performance of learning on the target may be reduced if irrelevant sources are included in transfer learning, which is defined as \textit{negative transfer} \citep{pan2009survey,ge2014handling}. It forces us to take the third issue into consideration: \textit{when to transfer}.

To avoid negative transfer, \citet{li2022trans} and \citet{lin2022transfer} proposed to aggregate a collection of candidate estimators, which leads to a relatively robust estimator. However, previous research on QR model aggregation has been restricted to the out-of-sample prediction error \citep{lu2015jackknife,wang2021jackknife}. Relatively little is understood about the optimality of the estimation error after model aggregation. 

Intuited by \citep{Eaton2008modeling}, we propose to select the sources with positive transfer by thresholding the change in performance on the target between learning with and without each source. Specifically, we randomly split the target index set $\mathcal{I}_0$ into the training part $\mathcal{I}_0^{\mathrm{tr}}$ and the validation part $\mathcal{I}_0^{\mathrm{va}}$ with equal size. We first obtain a benchmark estimate $\hat{\boldsymbol{\beta}}^{(0)}$ by fitting $\ell_{1}$-SQR on the target training data. Next, we carry out $\ell_{1}$-SQR based on the $k$-th source data as well as the target training data to obtain an estimate $\hat{\boldsymbol{\beta}}^{(k)}$, which is exactly the transferring step in Algorithm \ref{alg:otqr} with $\mathcal{A}_{\eta}=\{k\}$ and $\mathcal{I}_0=\mathcal{I}_0^{\mathrm{tr}}$. Then we examine the difference of quantile losses on the target validation data based on $\hat{\boldsymbol{\beta}}^{(k)}$ and $\hat{\boldsymbol{\beta}}^{(0)}$, denoted by $\widehat{T}^{(k)}=\widehat{Q}^{(0)}(\hat{\boldsymbol{\beta}}^{(k)};\mathcal{I}_0^{\mathrm{va}})-\widehat{Q}^{(0)}(\hat{\boldsymbol{\beta}}^{(0)};\mathcal{I}_0^{\mathrm{va}})$, where $\widehat{Q}^{(0)}(\boldsymbol{w};\mathcal{I})=1/|\mathcal{I}|\sum_{i\in\mathcal{I}}\rho_{\tau}(y_i^{(0)}-{({\boldsymbol{x}_i}^{(0)})}^{\top}{\boldsymbol{w}})$. We call $\widehat{T}^{(k)}$ the transferability index of the $k$-th study, which is preferable when it is smaller. The transferable set $\mathcal{A}_{\eta}$ is then estimated by the source datasets whose transferability indices are lower than a prespecified threshold. The detailed source detection procedure is presented in Algorithm \ref{alg:tqr}.

\begin{algorithm}[h]
	\caption{Trans-SQR Algorithm}
	\label{alg:tqr} 
	\begin{algorithmic}[1]
		\REQUIRE Target data $(\boldsymbol{X}^{(0)}, \boldsymbol{y}^{(0)})$, source data $\{(\boldsymbol{X}^{(k)}, \boldsymbol{y}^{(k)})\}_{k=1}^{K}$, penalty parameters $\{\lambda^{(k)}\}_{k=0}^{K}$,  bandwidths $\{h^{(k)}\}_{k=0}^{K}$ and a threshold $t$.
		\STATE\underline{\textbf{Estimating transferability index:}} 
		
		\noindent(1) Randomly split $\mathcal{I}_0=[n_0]$ into $\mathcal{I}_0=\mathcal{I}_0^{\mathrm{tr}}\cup\mathcal{I}_0^{\mathrm{va}}$ with $|\mathcal{I}_0^{\mathrm{va}}|=\lfloor n_0/2\rfloor$.
		
		\noindent(2) Compute
		$\hat{\boldsymbol{\beta}}^{(0)}\leftarrow\ell_1\text{-SQR}(\{\boldsymbol{x}_{i}^{(0)},{y}_{i}^{(0)}\}_{i\in\mathcal{I}_0^{\mathrm{tr}}};\lambda^{(0)},h^{(0)})$.
		
		\noindent(3) For $k=1,\ldots,K$, compute  $$\hat{\boldsymbol{\beta}}^{(k)}\leftarrow\ell_1\text{-SQR}(\{\boldsymbol{x}_{i}^{(k)},{y}_{i}^{(k)}\}_{i\in\mathcal{I}_k}\cup\{\boldsymbol{x}_{i}^{(0)},{y}_{i}^{(0)}\}_{i\in\mathcal{I}_0^{\mathrm{tr}}};\lambda^{(k)},h^{(k)}).$$
		
		\noindent(4) For $k=1,\ldots,K$, compute $\widehat{T}^{(k)}=\widehat{Q}^{(0)}(\hat{\boldsymbol{\beta}}^{(k)};\mathcal{I}_0^{\mathrm{va}})-\widehat{Q}^{(0)}(\hat{\boldsymbol{\beta}}^{(0)};\mathcal{I}_0^{\mathrm{va}})$.
		
		\STATE \underline{\textbf{Source detection:}} Let  $\widehat{\mathcal{A}}=\{k:\widehat{T}^{(k)}<t(\widehat{Q}^{(0)}(\hat{\boldsymbol{\beta}}^{(0)};\mathcal{I}_0^{\mathrm{va}})\vee 0.01)\}$.
		
		\STATE \underline{\textbf{Trans-SQR:}} $\hat{\boldsymbol{\beta}}\leftarrow$ run Algorithm \ref{alg:otqr} with $\{(\boldsymbol{X}^{(k)}, \boldsymbol{y}^{(k)})\}_{k\in\widehat{\mathcal{A}}\cup\{0\}}$
		
		\ENSURE $\hat{\boldsymbol{\beta}}$
	\end{algorithmic}
\end{algorithm}

Under regular conditions, the detected transferable set $\widehat{\mathcal{A}}$ will be shown to exactly identify the unknown ${\mathcal{A}}_{\eta}$ for some $\eta$ with high probability. Furthermore, in comparison to the model aggregation method proposed in \cite{li2022trans}, our method greatly eases the computation burden since we do not need to execute the oracle transferring procedure repeatedly based on a set of candidate estimates of $\mathcal{A}_{\eta}$.

\section{Theoretical Guarantees}
\label{sec:theory}
In this section, we will investigate the statistical properties of the proposed algorithms in Section \ref{sec:method}. We first establish high probability bounds of our QR transfer estimator in Algorithm \ref{alg:otqr} in Section \ref{subsec:estbound}. In Section \ref{subsec:detect}, we establish the consistency of the source detection procedure in Algorithm \ref{alg:tqr}.

\subsection{Estimation error of Oracle-Trans-SQR}
\label{subsec:estbound}
We begin by imposing some regularity conditions in smoothed quantile regression. Denote the covariance matrix of $\boldsymbol{x}^{(k)}$ as $\Sigma^{(k)}=\left(\sigma^{(k)}_{i j}\right)_{1 \leq i,j \leq p}$. 

\begin{condition}\label{cond:density}
	For $k=0,\ldots,K$ and some constant $b_0>0$,  ${\inf_k}f_{\epsilon^{(k)}\mid \boldsymbol{x}^{(k)}}(t)\geq f_l>0$ for all $|t|\leq b_0$ and  ${\sup_k}f_{\epsilon^{(k)}\mid \boldsymbol{x}^{(k)}}(0) \leq f_u<\infty$ almost surely. Besides, there exists a constant $l_0>0$  such that $\left|f_{\epsilon^{(k)}\mid \boldsymbol{x}^{(k)}}(u)-f_{\epsilon^{(k)}\mid \boldsymbol{x}^{(k)}}(v)\right| \leq l_0|u-v|$ for all $u, v \in \mathbb{R}$ almost surely.
\end{condition}
\begin{condition}\label{cond:kernel}
	The kernel function $K: \mathbb{R} \rightarrow[0, \infty)$ is symmetric around zero, and satisfies
	
	\noindent(a)$\int_{-\infty}^{\infty} K(u) \mathrm{d} u=1$ and $\int_{-\infty}^{\infty} u^2 K(u) \mathrm{d} u<\infty$; 
	
	\noindent (b) $\kappa_{\ell}=\int_{-\infty}^{\infty}|u|^{\ell} K(u) \mathrm{d} u<\infty$ for $\ell=1,2, \ldots$ and $\kappa_l=\min _{|u| \leq 1} K(u)>0$.
\end{condition}
\begin{condition}\label{cond:x}
	For $k=0,\ldots,K$, the  covariate vector $\boldsymbol{x}^{(k)}\in \mathbb{R}^p$ is sub-Gaussian: there exists some $v_1 \geq 1$ such that $\mathbb{P}\left(\left|\boldsymbol{u}^{\mathrm{T}} \boldsymbol{x}^{(k)}\right| \geq v_1\|\boldsymbol{u}\|_2 \cdot t\right) \leq 2 e^{-t^2 / 2}$ for all $\boldsymbol{u} \in \mathbb{R}^p$ and $t \geq 0$. In addition, $\mu_k=\sup_{\boldsymbol{u}\in\mathbb{S}^{p-1}}\mathbb{E}|\langle\boldsymbol{u},\boldsymbol{x}\rangle|^k<\infty$ for $k=1,\ldots,4$ and $0<\gamma_{p}\leq{\inf_k}\Lambda_{\mathrm{min}}\left({\Sigma}^{(k)}\right)\leq{\sup_k}\Lambda_{\mathrm{max} }\left({\Sigma}^{(k)}\right)\leq \gamma_1$. 
\end{condition}

\begin{remark}
	Condition \ref{cond:density} imposes regular conditions on the conditional density of the random errors. We allow for heteroscedastic errors. Condition \ref{cond:density}. Condition \ref{cond:kernel} is commonly used for kernel functions \citep{fernandes2021smoothing,tan2022conv}. Some detailed comments can be found in Remark 4.1 of \citet{tan2022conv}.  Condition \ref{cond:x} assumes sub-Gaussian covariates with well-behaved covariance structures. We comment that our theoretical analysis will still carry through if we impose a weaker restricted eigenvalue condition similar to that in \citet{bickel2009simultaneous}.
\end{remark}

The next condition characterizes the differences between $\boldsymbol{M}_{\mathcal{A}_{\eta}}$ and $\boldsymbol{M}_k$ defined in Section \ref{subsec:kqr}, which is used to ensure that
the common component $\boldsymbol{w}^{\mathcal{A}_{\eta}}$ to be transferred is close to our target $\boldsymbol{\beta}$.
\begin{condition}\label{cond:M1}
	Denote $\boldsymbol{M}_{\mathcal{A}_{\eta}}=\sum_{k \in \mathcal{A}_{\eta}\cup\{0\}} \alpha_{k} \boldsymbol{M}_k$. For some constant $c_M$, it holds that $\sup_{k\in\mathcal{A}_{\eta}}\|\boldsymbol{M}_{\mathcal{A}_{\eta}}^{-1}\boldsymbol{M}_k\|_1= c_M<\infty$.
\end{condition}
\begin{remark}
	Condition \ref{cond:M1} is related in spirit to Assumption 4 in \citet{tian2022transfer} as well as Condition 4 in \citet{li2022trans}.
	We regard Condition \ref{cond:M1} as reasonable as it does not require exact sparsity of $\boldsymbol{M}_{\mathcal{A}_{\eta}}^{-1}\boldsymbol{M}_k$. For example, the $\ell_1$-norm of the inverses are bounded for symmetric diagonally dominant positive matrices \citep{Li2008TheIN}
	as well as matrices with its $(i,j)$-th component being $\rho^{|i-j|}$ for some $\rho\in(-1,1)$. Besides, it also holds true for banded matrices with a fixed bandwidth \citep{demko1977inverses}.
\end{remark}

\begin{condition}\label{cond:xeta}
	Assume that  $\|\boldsymbol{x}^{(k)}\|_{\infty}\leq B_k$ almost surely for some $B_k>0$ and  $B_{k}\|{\boldsymbol{\delta}}_{\mathcal{A}}-{\boldsymbol{\delta}}^{(k)}\|_1\leq b_0$, for all $k\in\mathcal{A}_{\eta}\cup\{0\}$.
\end{condition}

\begin{remark}
	The upper bound $B_k$ in Condition \ref{cond:xeta} can be $O(1)$ when $\boldsymbol{x}^{(k)}$ has bounded components and $O(\log p)$ for general sub-Gaussian designs. Therefore, a sufficient condition for Condition \ref{cond:xeta} to hold is $\eta\lesssim 1$ in the former case and $\eta \lesssim 1/\log p$ in the latter.
\end{remark}

With the above preparations, we are ready to present the main result for our Oracle-Trans-SQR algorithm. Consider the parameter space
\begin{equation*}
\Theta(s, \eta)=\left\{\boldsymbol{B}=\left(\boldsymbol{\beta}, \boldsymbol{\delta^{(1)}}, \ldots, \boldsymbol{\delta}^{(K)}\right):\|\boldsymbol{\beta}\|_0 \leq s, \sup _{k \in \mathcal{A}_{\eta}}\left\|\delta^{(k)}\right\|_1 \leq \eta\right\}.
\end{equation*}

\begin{theorem}\label{thm:esterr}
	Assume Conditions \ref{cond:density}-\ref{cond:xeta} hold. Suppose that  $n_0\gtrsim \log p$, $n_{\mathcal{A}_{\eta}}\gtrsim s^2\log p $, and $\eta=O(1)$. Let $\lambda_{\boldsymbol{w}}\asymp{(\log p/(n_{\mathcal{A}_{\eta}}+n_0))}^{1/2}$ and $\lambda_{\boldsymbol{\delta}}\asymp{(\log p/n_0)}^{1/2}$. Choose the bandwidths as $h_{\boldsymbol{w}} \asymp{(s\log p / (n_{\mathcal{A}_{\eta}}+n_0))}^{1 / 4}$ and $h_{\boldsymbol{\delta}} \asymp{(\log p / n_0)}^{1 / 4}$. Then for the estimator $\hat{\boldsymbol{\beta}}$ obtained from the Oracle-Trans-SQR algorithm, there exist positive constants $c_1$ and $c_2$ such that 
	\begin{equation*}\label{eq:l1bound}
	\inf _{\boldsymbol{B} \in \Theta(s, {\eta})} \mathbb{P}\left(\|\hat{\boldsymbol{\beta}}-\boldsymbol{\beta}\|_1 \lesssim s\left(\frac{\log p}{n_{\mathcal{A}_{\eta}}+n_0}\right)^{1 / 2}+{\eta}\right) \geq 1- c_1\exp(-c_2\log p),
	\end{equation*}
	\begin{equation*}\label{eq:l2bound}
	\inf _{\boldsymbol{B} \in \Theta(s, \eta)} \mathbb{P}\left(\|\hat{\boldsymbol{\beta}}-\boldsymbol{\beta}\|_2 \lesssim\left(\frac{\log p}{n_0}\right)^{\frac{1}{4}}\left(\sqrt{s}{\left(\frac{\log p}{n_{\mathcal{A}_{\eta}}+n_0}\right)}^{\frac{1}{4}}+\sqrt{\eta}\right)\right) \geq 1- c_1\exp(-c_2\log p).
	\end{equation*}
\end{theorem}
\begin{remark}
	The above theorem established the convergence rate of $\hat{\boldsymbol{\beta}}$ from the Oracle-Trans-SQR algorithm under a proper choice of bandwidths. For single-task smoothed quantile regression based solely on the target data, \citet{tan2022conv} obtained the rate for $\ell_2$ bound as $\sqrt{s\log p/n_0}$. Hence, our proposed estimator will enjoy a sharper convergence rate provided that $\eta\ll s\sqrt{\log p/n_0}$ and $n_{\mathcal{A}} \gg n_0$. These results parallel analogous results in \citet{li2022trans} for linear regression and in \citet{tian2022transfer} for generalized linear models. 
\end{remark}

\subsection{Source detection consistency}
\label{subsec:detect}
Next, we investigate the theoretical property of the Trans-SQR algorithm. The key ingredient of this part is to establish the detection consistency of the transferable set $\mathcal{A}_{\eta}$. Let $\mathcal{A}_{\eta}^{c}=\{1,\ldots,K\}\setminus \mathcal{A}_{\eta}$ and $n_{\min}=\min_{1\leq k\leq K}n_k$. Similar to (\ref{eq:wAmoment}), let ${\boldsymbol{\beta}}^{(k)}$ be the parameter identified by 
\begin{equation*}\label{eq:bkmoment}
\sum_{k^{\prime} \in\{0,k\}}\alpha_{k^{\prime}}\mathbb{E}\left[\boldsymbol{x}^{(k^{\prime})}\left(F_{\epsilon^{(k^{\prime})}\mid \boldsymbol{x}^{(k^{\prime})}}\left({(\boldsymbol{x}^{(k^{\prime})})}^{\top}{(\boldsymbol{\beta}^{(k)}-\boldsymbol{w}^{(k^{\prime})})}\mid \boldsymbol{x}^{(k^{\prime})}\right)-\tau\right)\right]=0.
\end{equation*}
\begin{condition}\label{cond:sparsity}
	Suppose that there exist some positive $s^{\prime}$ and $\eta^{\prime}$ independent of $k$ such that ${\boldsymbol{\beta}}^{(k)}$ can be decomposed into $\boldsymbol{\beta}^{(k)}=\boldsymbol{\iota}^{(k)}+\boldsymbol{\varpi}^{(k)}$ with $\left\|\boldsymbol{\iota}^{(k)}\right\|_0 \leq s^{\prime}$ and $\left\|\boldsymbol{\varpi}^{(k)}\right\|_1 \leq {\eta}^{\prime}$ for $k=1,\ldots,K$. 
\end{condition}
\begin{remark}\label{remark:condsparsity}
	Condition \ref{cond:sparsity} imposes a ``weak'' sparse structure on ${\boldsymbol{\beta}}^{(k)}$, which is similar to Assumption 6 in \cite{tian2022transfer}. We allow the support of $\boldsymbol{\iota}^{(k)}$, denoted by $\mathcal{S}_k$, to be significantly different for $k\in {\mathcal{A}}_{\eta}$ and $k\in {\mathcal{A}}_{\eta}^{c}$.
\end{remark}

Define $\boldsymbol{R}^{(k)}=\mathbb{E}\left[\int_{0}^{1}\boldsymbol{x}^{(0)}{(\boldsymbol{x}^{(0)})}^{\top}f_{\epsilon^{(0)}\mid \boldsymbol{x}^{(0)}}(t{(\boldsymbol{x}^{(0)})}^{\top}(\boldsymbol{\beta}^{(k)}-\boldsymbol{\beta}))\mathrm{dt}\right]$. Let $\Omega_0=\sqrt{s\log p/n_0}$, $\Omega_{\mathrm{max}}=\sqrt{s^{\prime}\log p/(n_{\mathrm{min}}+n_0)}+(\log p/(n_{\mathrm{min}}+n_0))^{1/4}\sqrt{\eta^{\prime}}+(\log p/(n_{\mathrm{min}}+n_0))^{3/8}(s^{\prime})^{-1/8}\eta^{\prime}$, and $c_n=(\mu_1+2v_1\sqrt{\log p/n_0})(\Omega_{\mathrm{max}}+\Omega_0)$. 

\begin{condition}\label{cond:gap}
	Suppose that $\inf_{k\in{\mathcal{A}}_{\eta}^{c}}\Lambda_{\mathrm{min}}(\boldsymbol{R}^{(k)}):= \underline{\lambda}>0$,  $\sup_{k\in{\mathcal{A}}_{\eta}}\Lambda_{\mathrm{max}}(\boldsymbol{R}^{(k)}):= \bar{\lambda}<\infty$ and there exist some $\eta$ and a constant $C$ such that 
	$$\inf_{k\in{\mathcal{A}}_{\eta}^{c}}\|{\boldsymbol{\beta}}^{(k)}-{\boldsymbol{\beta}}\|_2^2\geq {\underline{\lambda}}^{-1}\left(\bar{\lambda}\sup_{k\in{\mathcal{A}}_{\eta}}\|{\boldsymbol{\beta}}^{(k)}-{\boldsymbol{\beta}}\|_2^2+12v_1\sqrt{\frac{\log p}{n_0}}\sup_{k\in[K]}\|{\boldsymbol{\beta}}^{(k)}-{\boldsymbol{\beta}}\|_2+2Cc_n\right).$$
\end{condition}
\begin{remark}
	Condition \ref{cond:gap} ensures the identifiability of some ${\mathcal{A}}_{\eta}$ by Trans-SQR. It assumes that for sources not in the transferable set ${\mathcal{A}}_{\eta}$, there is a significant gap between the population-level coefficient from the transferring step and the true coefficient of the target data. It is parallel to Assumption 5 in \cite{tian2022transfer}.
\end{remark}
\begin{theorem}\label{thm:detect}
	Assume that Conditions \ref{cond:density}-\ref{cond:sparsity} hold and Condition \ref{cond:gap} is satisfied for some $\eta$. Suppose that  $n_{\mathcal{A}_{\eta}}\gtrsim s^2\log p $. Let $\lambda_{\boldsymbol{w}}\asymp{(\log p/(n_{\mathcal{A}_{\eta}}+n_0))}^{1/2}$, and $\lambda_{\boldsymbol{\delta}}\asymp{(\log p/n_0)}^{1/2}$. For $k=0,\ldots,K$, let $\lambda^{(k)}\asymp{(\log p/(n_k+n_0))}^{1/2}$ and  $h^{(k)} \asymp{(s^{\prime}\log p / (n_k+n_0))}^{1 / 4}$. Choose the threshold $t$ as $\bar{\lambda} \sup_{k \in \mathcal{A}_{\eta}}\|{\boldsymbol{\beta}}^{(k)}-{\boldsymbol{\beta}}\|_2^2+\sqrt{\log p/n_0}\sup_{k \in [K]}\|{\boldsymbol{\beta}}^{(k)}-{\boldsymbol{\beta}}\|_2 + c_n \lesssim t \lesssim \underline{\lambda}/2 \inf_{k \in \mathcal{A}^{c}_{\eta}}\|{\boldsymbol{\beta}}^{(k)}-{\boldsymbol{\beta}}\|_2^2$. Then for the detected set $\widehat{\mathcal{A}}$ in Algorithm \ref{alg:tqr}, we have
	\begin{equation*}
	\mathbb{P}(\widehat{\mathcal{A}}=\mathcal{A}_{\eta})\rightarrow 1.
	\end{equation*}
	Consequently, the estimator $\hat{\boldsymbol{\beta}}$ obtained from the Trans-SQR algorithm enjoys the same $\ell_1/\ell_2$-estimation error upper bounds in Theorem \ref{thm:esterr} with high probability.
\end{theorem}
\begin{remark}
	Theorem \ref{thm:detect} indicates that the source detection procedure in Algorithm \ref{alg:tqr} can consistently detect the underlying transferable set $\mathcal{A}_{\eta}$.
\end{remark}

\section{Simulation}
\label{sec:simu}
In this section, we conduct numerical experiments to evaluate the empirical performance of the proposed transfer learning methods in comparison with some other alternatives. Specifically, we consider the following methods, including L1-SQR ($\ell_1$-SQR with only target data), Oracle-TSQR (Algorithm \ref{alg:otqr}), Oracle-TQR (Algorithm \ref{alg:otqr} with both the transferring and debiasing steps solved by $\ell_1$-QR), Naive-TSQR (Algorithm \ref{alg:otqr} with all sources) and TSQR (Algorithm \ref{alg:tqr}).  

\subsection{Data Generation}
We consider a target study with $n_0=150$ and $K=20$ source studies with $n_1,\ldots,n_K=100$. The dimension $p=500$ for both target and source data. We consider the following generative model for $k=0,\ldots,K$: 
\begin{equation*}
{y}^{(k)}=0.5+{(\boldsymbol{x}^{(k)})}^{\top}{\boldsymbol{w}}^{(k)}+\epsilon^{(k)},
\end{equation*}
where $\boldsymbol{x}^{(0)} \sim N_p(\mathbf{0}, \boldsymbol{\Sigma})$ with $\boldsymbol{\Sigma}=\left(0.7^{|i-j|}\right)_{1 \leq i,j \leq p}$ and $\boldsymbol{x}^{(k)}\sim\mathcal{N}\left(\mathbf{0}_p, \boldsymbol{\Sigma}+\boldsymbol{\epsilon} \boldsymbol{\epsilon}^T\right)$ with $\boldsymbol{\epsilon} \sim \mathcal{N}\left(\mathbf{0}_p, \delta^2 \boldsymbol{I}_p\right)$ for $k=1,\ldots,K$. We consider $\delta=0.3$ here. See \ref{subsec:simux} for additional simulations under more heterogeneous designs with larger $\delta$'s. We generate $\epsilon^{(k)}$ from two different distributions: (i) standard normal distribution, $\epsilon^{(k)}\sim \mathcal{N}(0,1)$; and (ii) $t$ distribution with degrees of freedom $3$, $\epsilon^{(k)}\sim t(3)$. 

Now we specify the coefficients in each study. For the target, we set $s=16$, $w^{(0)}_j=0.5\mathbf{I}\{j\in[s]\}$. Denote $r_{i}^{(k)}$ as independent Rademacher random variables (taking values in $\{1,-1\}$ with equal probability). For the source studies, we set 
\begin{equation*}
w_{j}^{(k)}=\begin{cases}w^{(0)}_{j}+ {\eta}/200*r_{j}^{(k)}\mathbf{I}\{j \in H^{(k)}\}, & k\in\mathcal{A}_{\eta},\\  {\eta}/100*r_{j}^{(k)}\mathbf{I}\{j \in H^{(k)}\cup [s]\}, & k\in\mathcal{A}_{\eta}^{c},\end{cases}
\end{equation*}
where $H^{(k)}$ is a random subset of $\{s+1,\ldots,p\}$ with $|H^{(k)}| = 200$.

We consider ${\eta} \in \{5,10,15\}$ for Gaussion errors and $\eta\in\{10,20,30\}$ for t errors. We vary $A=|\mathcal{A}_{\eta}|\in \{0,4,8,\ldots,20\}$ and repeat simulations $100$ times. For smoother QR, we use the Gaussian kernel for smoothing with regularization parameter selected by five-fold cross-validation (CV) and choose the bandwidths as $\max \{0.05, \sqrt{\tau(1-\tau)}\{\log (p) / n\}^{1 / 4}\}$ as recommended in \cite{tan2022conv} for a specific $n$ in the corresponding problem. It needs to be emphasized that this particular choice of bandwidths is by no means optimal numerically. One can also select the optimal bandwidths by CV. We have also noted that the performance of our proposed estimators is not sensitive to the choice of the bandwidths, see \ref{subsec:simuh} for additional simulation results about the numerical sensitivity to the bandwidths.

We set the threshold $t=0.2$ for the TQR estimator for simplicity although one can still choose $t$ by CV. We found that the performance of TQR is not sensitive to the choice of $t$ as long as $t$ falls into a proper interval. We also tried $t=0.05,0.1,0.15$ and the results are similar. 

For non-smoothed QR, the regularization parameter is selected by five-fold CV when $A=0$ and BIC when $A\geq 4$ respectively, since CV in $\ell_1$-QR greatly increases the computational burden as the sample size increases. BIC leads to similar performance as that of CV when the sample size is relatively large while spending much less time than CV. We evaluate the performance of each method in terms of the estimation error  $\|\hat{\boldsymbol{\beta}}-\boldsymbol{\beta}\|_2^2$ as well as the computation time.

\subsection{Simulation Results}
\begin{figure}[t!]
	\centering\includegraphics[width=4.5in]{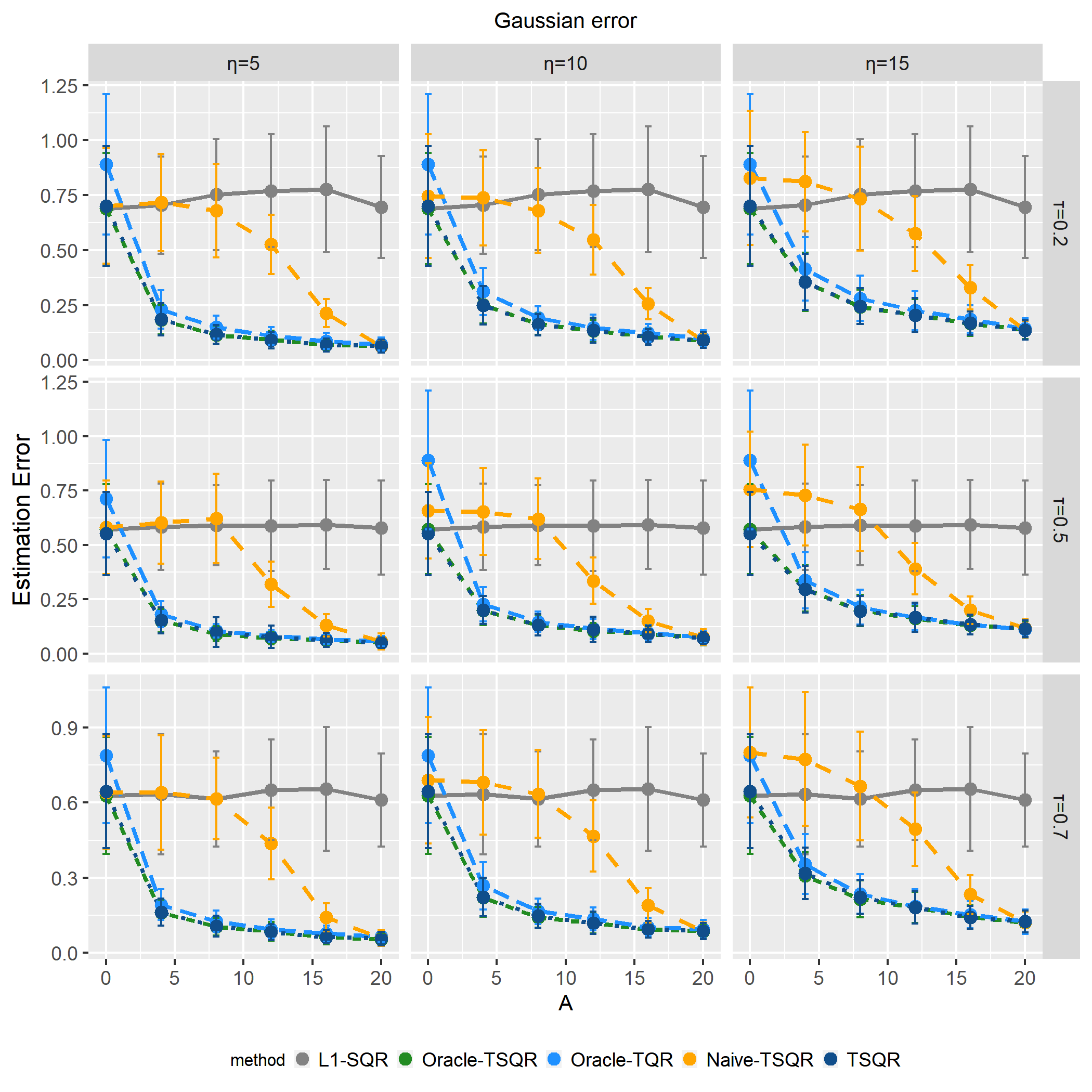}\par
	\caption{$\ell_2$ estimation errors of various methods with Gaussian errors under quantile levels $\tau=0.2, 0.5, 0.7$, averaged over 100 replications.\label{fig:ga}}
\end{figure}

\begin{figure}[t!]
	\centering\includegraphics[width=4.5in]{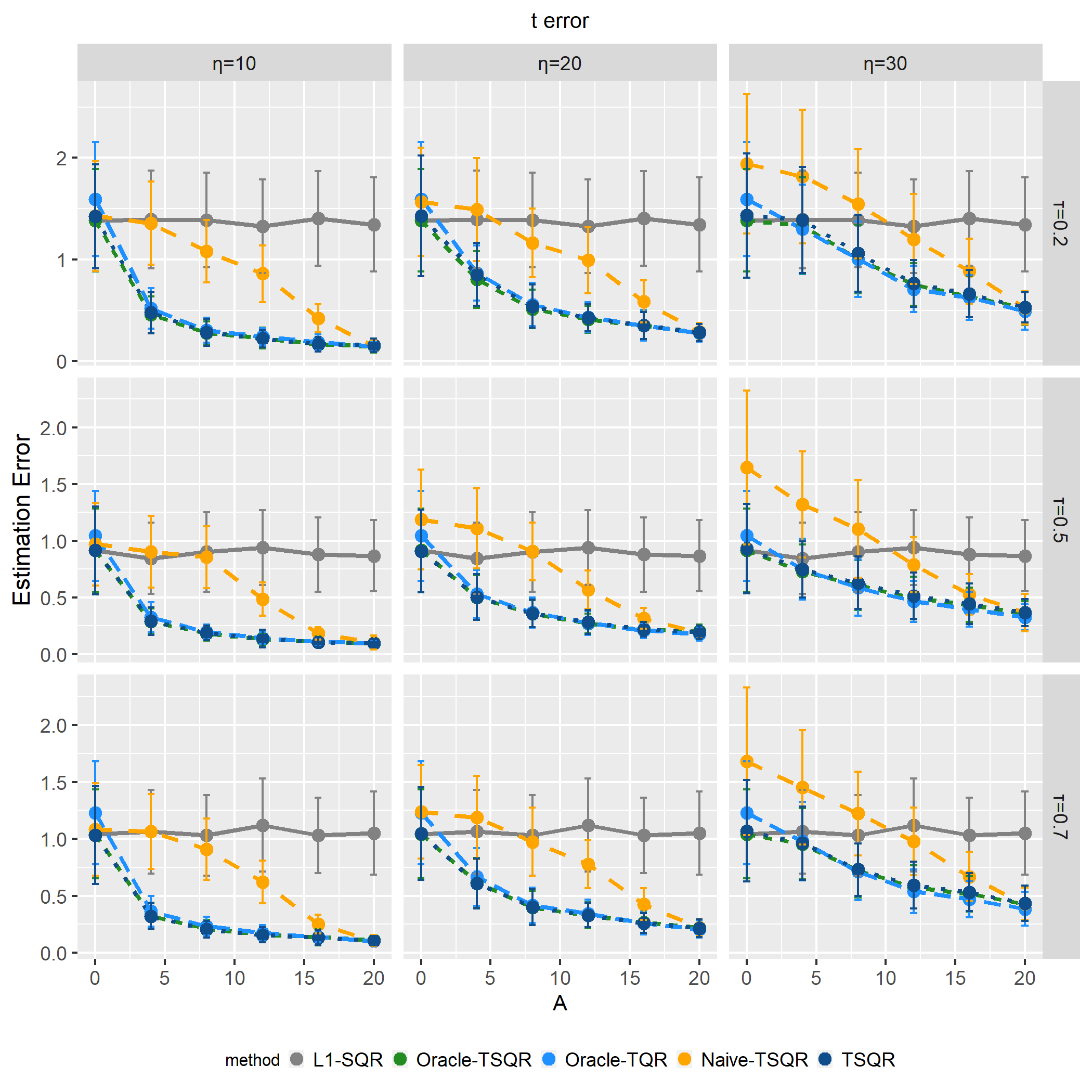}\par
	\caption{$\ell_2$ estimation errors of various methods with t errors under quantile levels $\tau=0.2, 0.5, 0.7$, averaged over 100 replications.\label{fig:t}}
\end{figure}

\begin{figure}[t!]
	\centering\includegraphics[width=3in]{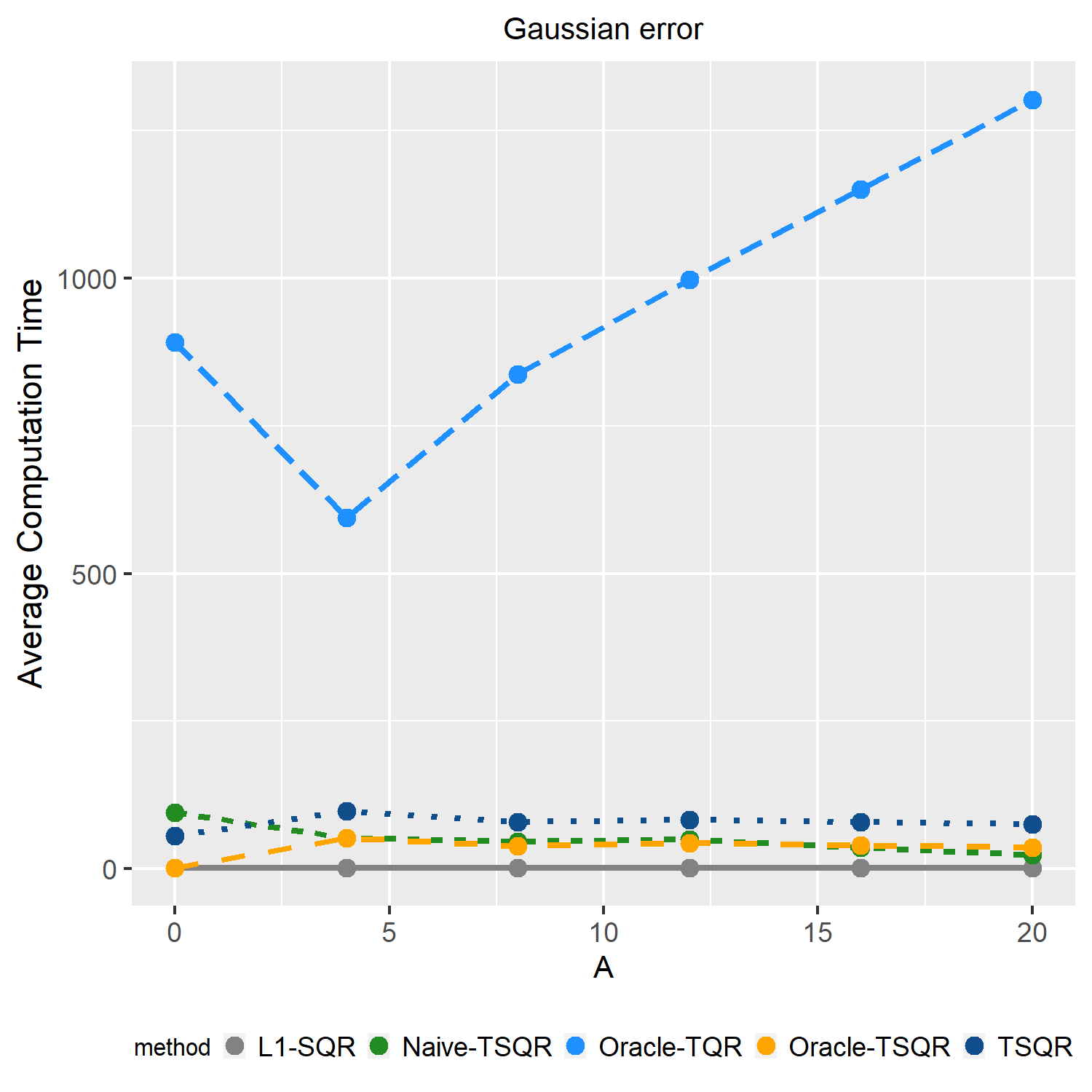}\par
	\caption{Computation time (in seconds) for each method and each $A$ under Gaussian errors, averaged over 100 replications and various $(\tau,h)$.\label{fig:time}}
\end{figure}

Figure \ref{fig:ga} and Figure \ref{fig:t} provide a comparison of $\ell_2$-estimation errors of the five methods mentioned above under Gaussian errors and t errors respectively. The average computation time of one replication for the five methods under Gaussian errors is reported in Figure \ref{fig:time}, where we take an average of over 100 replications and various $(\tau,h)$ for each method and each $A$. We highlight four conclusions in the following. 

Firstly, as we expect, Oracle-TSQR enjoys the best performance across all scenarios as it transfers knowledge exactly from the ideal transferable set $\mathcal{A}_{\eta}$. Moreover, smaller $\eta$ and bigger $A$ produce smaller estimation errors. In contrast, they do not affect the performance of the single-task $\ell_1$-SQR. This corroborates our main message that one can benefit from transfer learning, which is consistent with results in Theorem \ref{thm:esterr}.

Secondly, the worse performance of Naive-TSQR compared to the $\ell_1$-SQR when the size of the transferable set $A$ is small confirms our concern about the negative transfer. A larger $\eta$ leads to more severe negative transfer when $A$ is small and therefore, a larger $A$ is needed for Naive-TSQR to perform better than $\ell_1$-SQR. It also demonstrates the necessity of the source detection procedure in our Trans-SQR algorithm. 

Thirdly, our proposed TSQR almost nearly matches the oracle estimation errors obtained by Oracle-TSQR. This supports our theory on source detection consistency in Theorem \ref{thm:detect}. 

To illustrate the benefit of smoothing, we note in Figure \ref{fig:time} that oracle-TSQR and even TSQR have an overwhelming advantage in computational time compared to that of the non-smooth oracle-TQR especially when the sample size of transferable sources is large, even we have already used BIC to select the regularization parameter for oracle-TQR in these cases. In addition, we also note that Oracle-TSQR performs slightly better than Oracle-TQR in terms of the estimation error when the size of the transferable set $\mathcal{A}_{\eta}$ is relatively small. This phenomenon is also noted in \cite{tan2022conv} thanks to the strong convexity brought by convolution smoothing.

Generally speaking, the above conclusions on the performance of different methods apply when the errors follow a $t$ distribution.

\section{Application}
\label{sec:appl}

Primary familial brain calcification (PFBC) is a rare, genetically dominant, inherited neurological disorder characterized by bilateral calcifications in the basal ganglia and other brain regions, and commonly presents motor, psychiatric, and cognitive symptoms. Recent studies have identified JAM2 as a novel causative gene of autosomal recessive PFBC \citep{Cen2019jam2,Schottlaender2020jam2}. JAM2 encodes junctional adhesion molecule 2, which is highly expressed in neurovascular unit-related cell types (endothelial cells and astrocytes) and is predominantly localized on the plasma membrane. It may be important in cell-cell adhesion and maintaining homeostasis in the central nervous system (CNS). \cite{Schottlaender2020jam2} show that JAM2 variants lead to reduction of JAM2 mRNA expression and absence of JAM2 protein in patient's fibroblasts, consistent with a loss-of-function mechanism. Therefore, it is of great interest in predicting the expression level of gene JAM2 in target brain tissues, especially at the lower quantile levels.

We apply the proposed transfer learning algorithm to the Genotype-Tissue Expression (GTEx) data, available at https://gtexportal.org/. This dataset contains gene expression levels from 49 tissues of 838 individuals, a total of 1,207,976 observations of 38,187 genes. We study the CNS gene regulations in different tissues. The CNS-related genes were assembled as MODULE\_137, which includes 545 genes in total as well as 1,632 additional genes that are significantly enriched in the same experiments as the genes of the module (see http://robotics.stanford.edu/~erans/cancer/modules/module\_137 for a detailed description of this module).

Specifically, we are interested in predicting the expression level of gene JAM2 in a target tissue using other CNS genes. We consider 13 brain tissues as our target tissues. The corresponding models are estimated one by one. The average sample sizes of the target dataset and source datasets are 177 and 14837 respectively. For each model, we split the target dataset into five folds with each fold being predicted using the remaining four folds as the training data. The regularization parameters are selected by five-fold cross-validation. 

\begin{figure}[t!]
	\centering\includegraphics[width=4.5in]{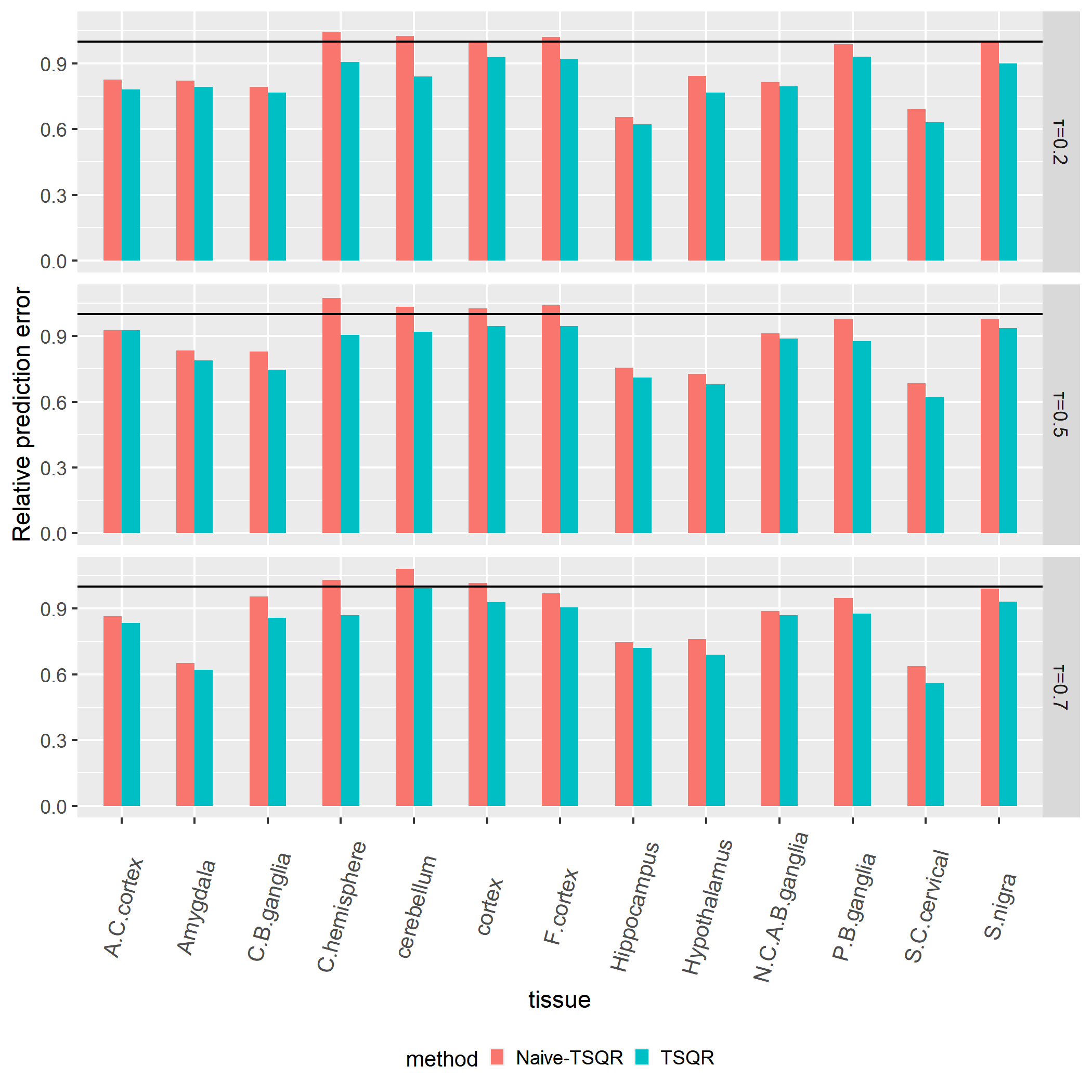}\par
	\caption{Prediction	errors of the Naive-TSQR and TSQR relative to the L1-SQR for the expression level of gene JAM2 in 13 brain tissues under quantile levels $0,2$, $0.5$ and $0.7$, evaluated via fivefold cross-validation. The black horizontal line here represents ratio 1.\label{fig:jam2}}
\end{figure}

Figure \ref{fig:jam2} reports the relative prediction errors of the Naive-TSQR and our proposed TSQR to the $\ell_1$-SQR estimator based solely on the target data. As we expected, transferring-based methods reduce prediction errors in most cases. In addition, the Naive-TSQR without the source detection procedure performs worse than L1-SQR on the Hemisphere and Cerebellum tissues, while our proposed TSQR reflects its robustness to avoid including unrelated sources across all scenarios. In the best scenario, TSQR reduces the prediction error by about 40\%, see the prediction of the JAM2 expression at the $0.7$-th quantile on the S.C.cervical tissue for example.

Another interesting finding is that in comparison to the analogous empirical analysis results in \cite{li2022trans}, where transfer learning improves the prediction error by about 30\% in the mean regression world for the Hippocampus tissue, here in the quantile regression world, we find that the inclusion of sources seems to be less useful at quantile levels 0.5 and 0.7 but becomes quietly helpful at the lower quantile level 0.2. This indicates that the similarity between the target and the sources may vary with the distribution of our responses. Since the expression level of JAM2 at the lower quantile is of more interest to us, our QR model for transfer learning is more flexible than that of the mean regression.
\section{Discussion}
\label{sec:disc}
In this article, we study the high-dimensional quantile regression problem under the transfer learning setting. We propose a two-step framework for QR transfer via convolution smoothing as well as a clustering-based source detection procedure to avoid negative transfer. Numerical experiments as well as an empirical study about GTEx data validate the effectiveness of our proposed method.

There are several avenues for future research. Firstly, there may exist subgroup structures in the target dataset, which may be exploited for precise transfer learning. Specifically, we may consider the target parameter $\boldsymbol{\beta}_{g}$ with $g\in\{ 1, \ldots, G\}$ indicating the subgroup and borrow information from other sources with a similar group structure. Secondly, transfer learning of datasets with more complex structures, like matrices or tensors may be of interest as well.

\appendix
\section{Additional Simulation results}\label{sec:addsimu}
\subsection{QR Transfer with More Heterogeneous Designs}\label{subsec:simux}
\begin{figure}[t!]
	\centering\includegraphics[width=5in]{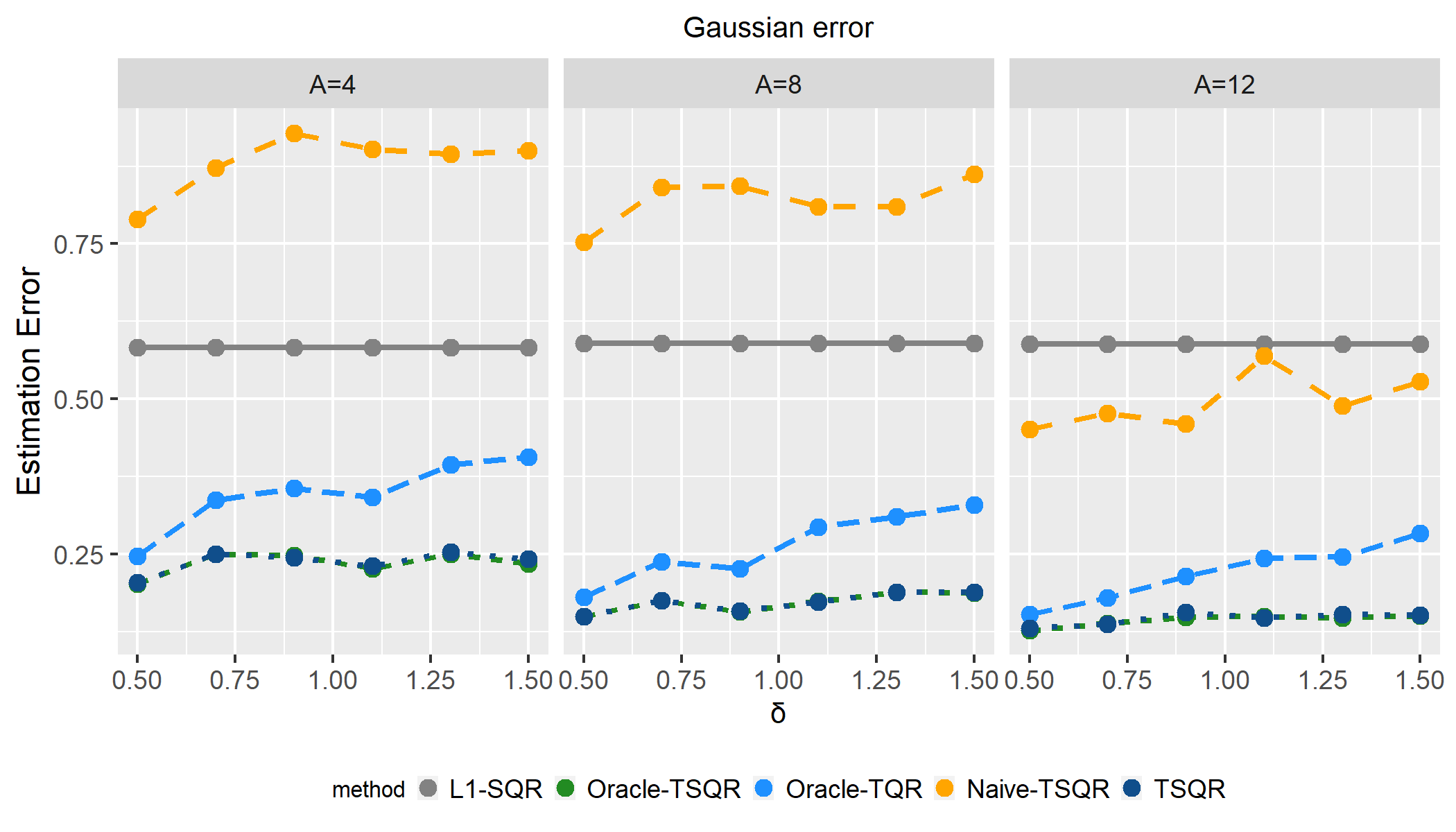}\par
	\caption{$\ell_2$ estimation errors of various methods with respect to $\delta$ under Gaussian errors at quantile level $\tau=0.5$ and $\eta=10$, averaged over 100 replications. Here the horizontal axis $\delta$ represents the heterogeneity parameter of the designs. \label{fig:ga_x}}
\end{figure}

In this subsection, we conduct additional simulations to investigate the impact of heterogeneous designs on the performance of the transferred estimators. Specifically, we consider the same setting as that in Section 4 of the main paper under Gaussian errors with the only difference in the generation of covariates in the sources. We consider $\boldsymbol{x}^{(0)} \sim N_p(\mathbf{0}, \boldsymbol{\Sigma})$ with $\boldsymbol{\Sigma}=\left(0.7^{|i-j|}\right)_{1 \leq i,j \leq p}$ and $\boldsymbol{x}^{(k)}\sim\mathcal{N}\left(\mathbf{0}_p, \boldsymbol{\Sigma}+\boldsymbol{\epsilon} \boldsymbol{\epsilon}^T\right)$ with $\boldsymbol{\epsilon} \sim \mathcal{N}\left(\mathbf{0}_p, \delta^2 \boldsymbol{I}_p\right)$ for $k=1,\ldots,K$. We fix $h=10$, $\tau=0.5$ and consider $A\in\{4,8,12\}$. We vary $\delta$ from $0.5$ to $1.5$ with stepsize $0.2$ and evaluate the change of performance with respect to $\delta$. The average estimation errors of the five methods ($\ell_1$-SQR, Oracle-TSQR, Oracle-TQR, Naive-TSQR, and TSQR) over 100 replications are displayed in Figure \ref{fig:ga_x}.

As we can see from Figure \ref{fig:ga_x}, the performance of our Oracle-TSQR as well as the TSQR remains stable as the heterogeneity parameter $\delta$ of the designs increases, while the performance of the Naive-TSQR and the non-smoothed Oracle-TQR becomes slightly worse as $\delta$ increases. This illustrates the stability of our smoothed QR transferring algorithms.

\subsection{Sensitivity to the smoothing bandwidths}\label{subsec:simuh}

\begin{figure}[t!]
	\centering\includegraphics[width=5in]{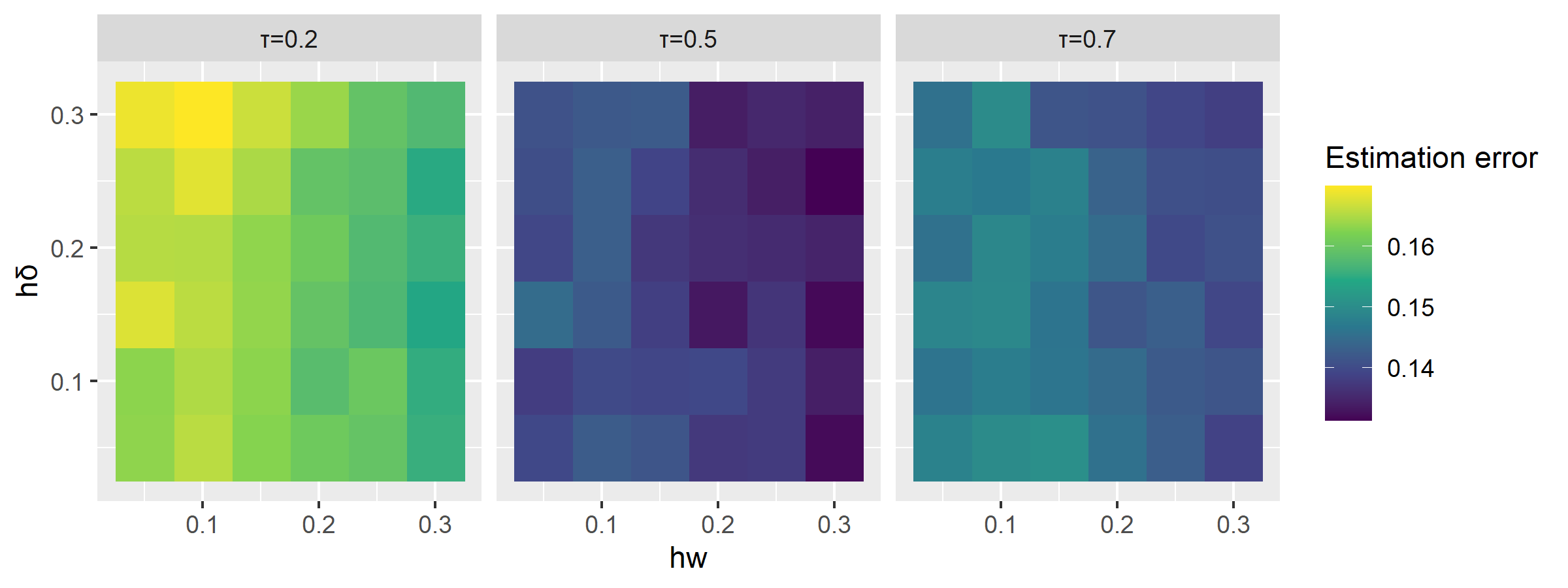}\par
	\caption{$\ell_2$ estimation errors of Oracle-TSQR with different choice of bandwidths under Gaussian errors at $\eta=10$, $A=8$, and quantile levels $\tau=0.2,0.5,0.7$. \label{fig:ga_h}}
\end{figure}

We investigate the sensitivity of our Oracle-TSQR to the smoothing bandwidths in this subsection. Specifically, we consider the same setting as that in Section 4 of the main paper under Gaussian errors with $\eta$ fixed at $10$ and $A$ fixed at $8$. For simulations in the main text, we choose the bandwidths as $\max \{0.05, \sqrt{\tau(1-\tau)}\{\log (p) / n\}^{1 / 4}\}$ as recommended in \cite{tan2022conv} for a specific $n$ in the corresponding problem. For example, for $A=8$, we have $h_{\boldsymbol{w}}\approx0.14$ and $h_{\boldsymbol{\delta}}\approx0.07$ at the quantile level $\tau=0.5$. Here we consider different choices of bandwidths $(h_{\boldsymbol{w}},h_{\boldsymbol{\delta}})$, both of which take values in $\{0.05,0.10,...,0.3\}$. There are 36 different combinations in total. For each combination choice of bandwidths, we replicate the simulation 100 times, and here in Figure \ref{fig:ga_h} we present the average $\ell_2$-estimation errors for each combination. For better comparison, we also report the average estimation errors of the five methods ($\ell_1$-SQR, Oracle-TQR, Oracle-TSQR, Naive-TSQR, and TSQR) with the smoother ones using bandwidths $\max \{0.05, \sqrt{\tau(1-\tau)}\{\log (p) / n\}^{1 / 4}\}$ at $\eta=10$
and $A=8$ under different quantile levels.

As we can see from Figure \ref{fig:ga_h}, the estimation errors are not very sensitive to the choice of the smoothing bandwidths $h_{\boldsymbol{w}}$ and $h_{\boldsymbol{\delta}}$ under each quantile level. Let us take $\tau=0.5$ for illustration. As reported in Table \ref{tab:rec_h}, the average estimation error of Oracle-TSQR chosen by the recommended bandwidths in \cite{tan2022conv} is 0.1307 with a standard deviation of 0.0471. We can see from the middle panel that the estimation errors under various choices of $h_{\boldsymbol{w}}$ and $h_{\boldsymbol{\delta}}$ fall between 0.1314 and 0.1449, the range of which is approximately 1/3 of its standard deviation (0.0471). This suggests that the performance of our proposed Oracle-TSQR is not really sensitive to the smoothing bandwidths. We also note that under various choices of bandwidths, the estimation errors of our Oracle-TSQR all perform slightly better under the non-smoothed Oracle-TQR, which is 0.1466. This again illustrates the benefit of our convolution smoothing.

\begin{table}[t!]
	\caption{Estimation errors and standard deviations of various methods at $\eta=10$ and $A=8$ under different quantile levels, with the smoothed estimators based on the recommended bandwidths.\label{tab:rec_h}}
	\begin{tabular}{cccccc}
		\hline\hline
		& L1-SQR          & Oracle-TSQR        & Oracle-TQR         & Naive-TSQR        & TSQR            \\\hline
		$\tau$& \multicolumn{5}{c}{$\ell_2$-estimation error (standard deviation)}
		\\\hline
		$0.2 $   & 0.7530 (0.2530) & \textbf{0.1631 (0.0479)} & 0.1918 (0.0532) & 0.6804 (0.1935) & 0.1609 (0.0488) \\\hline
		$0.5 $     & 0.5897 (0.1840) & \textbf{0.1307 (0.0471)} & 0.1466 (0.0471) & 0.6202 (0.1859) & 0.1322 (0.0485) \\\hline
		$0.7$    & 0.6153 (0.1903) & \textbf{0.1406 (0.0408)} & 0.1648 (0.0516) & 0.6353 (0.1767) & 0.1462 (0.0482)\\\hline\hline
	\end{tabular}
\end{table}

Similar conclusions also apply for the quantile levels $\tau=0.2$ and $\tau=0.7$ and thus omitted here. In conclusion, we recommend to choose the bandwidths as that in \cite{tan2022conv}, although in practice one can also use CV for selecting the bandwidths for optimal numeric performance.

\section{Distributed QR Transfer}\label{sec:distsqr}
\subsection{Distributed QR Transfer Algorithm}
Here we adopt the approximate Newton-type method proposed by \cite{shamir2014communication} and further examined in \cite{jordan2018communication} and \cite{wang2017efficient} to solve the transferring step in Algorithm \ref{alg:otqr} in a distributed manner.

With our loss of generality, we set $\mathcal{A}_\eta=\{1,\ldots,|\mathcal{A}_{\eta}|\}$. Let $\boldsymbol{\alpha}$ denote a $|\mathcal{A}_{\eta}|+1$-dimensional vector with the $k+1$-th element being $\alpha_k$. We first generate the pilot sample sizes $\{n^{*}_k\}_{k\in\mathcal{A}\cup\{0\}}$ from multinomial distribution $\mathcal{M}(n_{*},\boldsymbol{\alpha})$ with $n_{*}=\rho_0 (n_{\mathcal{A}_{\eta}}+n_0)$ for some $\rho_0\in(0,1)$. For each $k\in\mathcal{A}\cup\{0\}$, we randomly select $n^{*}_{k}$ samples from the $k$-th site, with the index set denoted by $\mathcal{D}^{*}_{k}$. Transfer $\{((\boldsymbol{x}^{(k)}_i)^{\top},y_i^{(k)})\}_{i\in\mathcal{D}^{*}_k}$ from the $k$-th site to the target site.

Denote the empirical smoothed quantile loss on the pilot pooled data by  $$\widehat{Q}_{h_{*}}^{*}(\boldsymbol{w})=\frac{1}{n_*h_{*}}\sum_{k\in\mathcal{A}_{\eta}\cup\{0\}}\sum_{i\in \mathcal{D}^{*}_k}\int_{-\infty}^{\infty}\rho_{\tau}(u)K\left(\frac{u+{\boldsymbol{w}}^{\top}\boldsymbol{x}_i^{(k)}-y_i^{(k)}}{h_{*}}\right)\mathrm{d}u$$. The gradient vectors of $\widehat{Q}^{*}_{h_{*}}(\boldsymbol{w})$ are given respectively by
$$
\nabla\widehat{Q}^{*}_{h_{*}}(\boldsymbol{w})=1/n^{*}\sum_{k\in\mathcal{A}\cup\{0\}}\sum_{i\in\mathcal{D}^{*}_k}\{\bar{K}(({(\boldsymbol{x}_i^{(k)})}^{\top}\boldsymbol{w}-y_i^{(k)})/h_{*})-\tau\}\boldsymbol{x}_i^{(k)}.
$$

Given an initial estimator $\tilde{\boldsymbol{w}}^{(0)}$, consider the first-order Taylor expansion of $\widehat{Q}^{\mathcal{A}_{\eta}}_{h_{\boldsymbol{w}}}(\boldsymbol{w})$ around $\tilde{\boldsymbol{w}}^{(0)}$:
\begin{equation}\label{eq:ra}
\widehat{Q}^{\mathcal{A}_{\eta}}_{h_{\boldsymbol{w}}}(\boldsymbol{w})=\widehat{Q}^{\mathcal{A}_{\eta}}_{h_{\boldsymbol{w}}}( \tilde{\boldsymbol{w}}^{(0)})+\langle\nabla\widehat{Q}^{\mathcal{A}_{\eta}}_{h_{\boldsymbol{w}}}(\tilde{\boldsymbol{w}}^{(0)}),\boldsymbol{w}-\tilde{\boldsymbol{w}}^{(0)}\rangle+R^{\mathcal{A}_{\eta}}(\boldsymbol{w}),
\end{equation}
where $R^{\mathcal{A}_{\eta}}(\boldsymbol{w})$ is the linear approximation error. In the distributed environment, the gradient $\nabla\widehat{Q}^{\mathcal{A}_{\eta}}_{h_{\boldsymbol{w}}}(\tilde{\boldsymbol{w}}^{(0)})$ can be easily be communicated. Therefore, it suffices to find a good replacement of $R^{\mathcal{A}_{\eta}}(\boldsymbol{w})$. Here we propose to use the analogous approximation error from the loss of the pilot pooled sample, that is, we approximate $R^{\mathcal{A}_{\eta}}(\boldsymbol{w})$ by 
\begin{equation}\label{eq:r*}
R^{\mathcal{A}_{\eta}}(\boldsymbol{w})\approx R^{*}(\boldsymbol{w})=\widehat{Q}^{*}_{h_{*}}(\boldsymbol{w})-\widehat{Q}^{*}_{h_{*}}( \tilde{\boldsymbol{w}}^{(0)})-\langle\nabla\widehat{Q}^{*}_{h_{*}}(\tilde{\boldsymbol{w}}^{(0)}),\boldsymbol{w}-\tilde{\boldsymbol{w}}^{(0)}\rangle.
\end{equation}
Plugging (\ref{eq:r*}) into (\ref{eq:ra})  motivates us to consider the surrogate smoothed quantile loss:
\begin{equation*}\label{eq:shiftloss}
\tilde{Q}(\boldsymbol{w})=\widehat{Q}^{*}_{h_{*}}({\boldsymbol{w}})-\langle\nabla\widehat{Q}^{*}_{h_{*}}(\tilde{\boldsymbol{w}}^{(0)})-\nabla\widehat{Q}^{\mathcal{A}_{\eta}}_{h_{\boldsymbol{w}}}(\tilde{\boldsymbol{w}}^{(0)}),\boldsymbol{w}\rangle.
\end{equation*}
Consequently, a communication-efficient penalized estimator for $\boldsymbol{w}^{\mathcal{A}_{\eta}}$ can be obtained by solving
\begin{equation}\label{eq:w1}
\tilde{\boldsymbol{w}}^{(1)}\in\underset{\boldsymbol{w}\in\mathbb{R}^{p}}{\arg\min}\quad \tilde{Q}(\boldsymbol{w})+\lambda_1\|\boldsymbol{w}\|_1.
\end{equation}
The above procedure could be done iteratively. In fact, we can define shifted losses $\tilde{Q}^{(t)}(\boldsymbol{w})=\widehat{Q}^{*}_{h_{*}}({\boldsymbol{w}})-\langle\nabla\widehat{Q}^{*}_{h_{*}}(\tilde{\boldsymbol{w}}^{(t-1)})-\nabla\widehat{Q}^{\mathcal{A}_{\eta}}_{h_{\boldsymbol{w}}}(\tilde{\boldsymbol{w}}^{(t-1)}),\boldsymbol{w}\rangle$ and obtain a sequence of estimators by solving $\tilde{\boldsymbol{w}}^{(t)}\in\underset{\boldsymbol{w}\in\mathbb{R}^{p}}{\arg\min} \tilde{Q}^{(t)}(\boldsymbol{w})+\lambda_t\|\boldsymbol{w}\|_1$ for $t=2,\ldots,T$. The details for our distributed Trans-SQR are presented in Algorithm \ref{alg:dist}.
\begin{algorithm}[h]
	\caption{Distributed-Oracle-Trans-SQR Algorithm}
	\label{alg:dist} 
	\begin{algorithmic}[1]
		\REQUIRE Target data $(\boldsymbol{X}^{(0)}, \boldsymbol{y}^{(0)})$, source data $\{(\boldsymbol{X}^{(k)}, \boldsymbol{y}^{(k)})\}_{k \in \mathcal{A}_{\eta}}$, pilot pooled sample $\{\{(\boldsymbol{x}^{(k)}_i,y_i^{(k)})\}_{i\in\mathcal{D}^{*}_k}\}_{k \in \mathcal{A}_{\eta}\cup\{0
			\}}$, an initial estimate $\tilde{\boldsymbol{w}}^{(0)}$, number of iterations $T$, penalty parameters $(\lambda_{\boldsymbol{w}},\lambda_{\boldsymbol{\delta}},\{\lambda_t\}_{t=1}^{T})$ and bandwidths 
		$(h_{\boldsymbol{w}},h_{\boldsymbol{\delta}},h_*)$.
		
		\STATE \underline{\textbf{Distributed Transferring:}} For $t=1,\ldots,T$, compute
		\begin{equation}\label{eq:transferringT}
		\tilde{\boldsymbol{w}}^{(t)}\leftarrow \underset{\boldsymbol{w}\in\mathbb{R}^{p}}{\arg\min}\quad\tilde{Q}^{(t)}(\boldsymbol{w})+\lambda_t\|\boldsymbol{w}\|_1
		\end{equation}
		
		\STATE \underline{\textbf{Debiasing:}} Compute
		\begin{equation*}\label{eq:debiasingT}
		\tilde{\boldsymbol{\delta}}^{(T)}\leftarrow \ell_{1}\text{-SQR}(\{(\boldsymbol{x}^{(0)}_{i},y^{(0)}_i-{(\boldsymbol{x}^{(0)}_{i})}^{\top}\tilde{\boldsymbol{w}}^{(T)})\}_{i=1}^{n_0};\lambda_{\boldsymbol{\delta}},h_{\boldsymbol{\delta}})
		\end{equation*}
		
		\ENSURE
		\begin{equation*}\label{eq:outputT}
		{\tilde{\boldsymbol\beta}}^{(T)}={\tilde{\boldsymbol{w}}}^{(T)}+\tilde{\boldsymbol\delta}^{(T)} 
		\end{equation*} 
	\end{algorithmic}
\end{algorithm}

As we will show in Theorem \ref{thm:distesterr}, a ``good'' estimator $\tilde{\boldsymbol{w}}^{(0)}$ is needed to guarantee the theoretical properties of $\tilde{\boldsymbol{w}}^{(T)}$. Taking the heterogeneity among the target and sources into consideration, we propose to obtain $\tilde{\boldsymbol{w}}^{(0)}$ on the pilot pooled sample by solving
\begin{equation}\label{eq:w0}
\tilde{\boldsymbol{w}}^{(0)}\in\underset{\boldsymbol{w}\in\mathbb{R}^{p}}{\arg\min} \quad\widehat{Q}^{*}_{h_*}(\boldsymbol{w})+\lambda_*\|\boldsymbol{w}\|_1.
\end{equation}
The optimization problem in (\ref{eq:transferringT}) could be solved by the local adaptive majorize-minimize (LAMM) algorithm \citep{fan2018lamm,tan2022communication}.

\subsection{Theory for Distributed-Oracle-Trans-SQR}\label{subsec:distestbound}
Here we establish the analogous estimation error bounds for our distributed QR transfer estimator. In addition to Condition \ref{cond:x}, we impose the following boundedness condition on the covariate vectors.
\begin{condition}\label{cond:xb}
	There exists some constant $B \geq 1$ such that $\max_{j\in[p]} |x^{(k)}_j| \leq B$ almost surely for all $k=0,\ldots,K$. 
\end{condition}

To better understand the mechanism of the distributed QR transfer estimator, we first present a deterministic result based on some ``good'' events. Define the $\ell_{2}$-ball $\mathbb{B}_2(r)=\{\boldsymbol{\delta}\in\mathbb{R}^{p}:\|\boldsymbol{\delta}\|_2\leq r\}$ and the cone-like set $\Lambda=\Lambda(s,\eta)=\{\boldsymbol{u}\in\mathbb{R}^{p}:\|\boldsymbol{u}\|_1\leq 5\sqrt{s}\|\boldsymbol{u}\|_2+4c_M\eta\}$. Consider the events
$
\mathcal{E}_0(r)=\{\tilde{\boldsymbol{w}}^{(0)}:\tilde{\boldsymbol{w}}^{(0)}-\boldsymbol{w}^{\mathcal{A}_{\eta}}\in\mathbb{B}_2(r)\cap\Lambda \}$ and $
\mathcal{E}_{\boldsymbol{w}}(\lambda_{\boldsymbol {w}})=\{\lambda_{\boldsymbol{w}}\geq2\|\nabla\widehat{Q}^{\mathcal{A}_{\eta}}_{h_{\boldsymbol{w}}}({\boldsymbol{w}}^{\mathcal{A}_{\eta}})-\nabla{Q}^{\mathcal{A}_{\eta}}_{h_{\boldsymbol{w}}}(\boldsymbol{w}^{\mathcal{A}_{\eta}})\|_{\infty}\}.$

\begin{proposition}\label{pro:disterr}
	Assume Conditions \ref{cond:density}-\ref{cond:xeta} and \ref{cond:xb} hold. Define $\rho_*=\sqrt{\log p/(n_{\mathcal{A}_{\eta}}+n_0)}/h_{\boldsymbol{w}}+\sqrt{\log p/n_*}/h_{*}$ for some $h_*>0$. Let $0<h_{\boldsymbol{w}}\leq h_*\lesssim 1$ and  $\lambda_1=\lambda_{\boldsymbol {w}}+\rho$ satisfy $\rho\asymp\max\{s^{-1/2}(h^2_{\boldsymbol{w}}+h_*r_*),r_{*}\sqrt{s}\rho_{*}+\eta\rho_{*}\}$ and $h_{*}\gtrsim \sqrt{\log p/(n_{*}h_{*})}\eta +\lambda_1\sqrt{s}+\sqrt{\lambda_1\eta}$. Then conditioning on the event $\mathcal{E}_0(r_{*})\cap\mathcal{E}_{\boldsymbol{w}}(\lambda_{\boldsymbol {w}})$, the one-step estimator $\tilde{\boldsymbol{w}}^{(1)}$ obtained by (\ref{eq:w1}) satisfies $\tilde{\boldsymbol{w}}^{(1)}\in\Lambda$ and
	\begin{equation}\label{eq:l2boundT1}
	\begin{aligned}
	&\inf _{\boldsymbol{B} \in \Theta(s, {\eta})} \mathbb{P}\left(\|\tilde{\boldsymbol{w}}^{(1)}-\boldsymbol{w}^{\mathcal{A}_{\eta}}\|_2\lesssim\lambda_{\boldsymbol {w}}\sqrt{s} + h_{\boldsymbol{w}}^2 + \varphi_0 r_* + \varphi_1\sqrt{\eta} + \varphi_2\eta\right)\\&\geq 1-c_1\exp(-c_2\log p),
	\end{aligned}
	\end{equation}
	where $\varphi_0=s\rho_*+h_{*}$, $\varphi_1=(s^{-1/4}\sqrt{h_*}+s^{1/4}\sqrt{\rho_*})\sqrt{r_*}+s^{-1/4}h_{\boldsymbol{w}}+\sqrt{\lambda_{\boldsymbol {w}}}$ and $\varphi_2=\sqrt{s}\rho_*+\sqrt{\rho_*}+\sqrt{\log p/(n_*h_*)}$.
\end{proposition}

The upper bound in (\ref{eq:l2boundT1}) can be decomposed into three parts: (i) the first two terms $\lambda_{\boldsymbol {w}}\sqrt{s} + h_{\boldsymbol{w}}^2$ is the nearly optimal rate when all transferable sources are used and there's no heterogeneity among the sources and target; (ii) the third term $\varphi_0 r_*$ is a contraction of the initial estimation error given by $\tilde{\boldsymbol{w}}^{(0)}$ and (iii) the last two terms could be seen as the price we pay for the heterogeneity among used data sets. 

With large enough samples in the pilot and pooled sources, that is, $n_*\gtrsim s^2\log p$ and $n_{\mathcal{A}_{\eta}}+n_0\gtrsim s^3\log p$, the contraction factor $\varphi_0$ can be strictly less than $1$, which will consequently improve the convergence rate of $\tilde{\boldsymbol{w}}^{(0)}$. 

When homogeneity is assumed among the sources and target—namely, $\eta=0$, Proposition \ref{pro:disterr} degenerates to Theorem 11 in \citet{tan2022communication}. Therefore, it can be seen as an extension of the established results for distributed smoothing QR estimators in \cite{tan2022communication} to allow for the existence of heterogeneity in the high-dimensional setting.

We are now ready to present the estimation error bounds for $\tilde{\boldsymbol{\beta}}^{(T)}$from Algorithm \ref{alg:dist}.

\begin{theorem}\label{thm:distesterr}
	Assume Conditions \ref{cond:density}-\ref{cond:xeta} and \ref{cond:xb} hold. Suppose that $n_*\gtrsim s^2\log p$, $n_{\mathcal{A}_{\eta}}\gtrsim s^3\log p$, $\eta\lesssim s\sqrt{\log p/n_*}$ and $\eta\lesssim(s^5\log p/n_*)^{1/8}$. Choose the regularization parameters $(\lambda_{\boldsymbol {w}},\lambda_{\boldsymbol{\delta}})$ and bandwidths $(h_{\boldsymbol{w}},h_{\boldsymbol{\delta}})$ as that in Theorem \ref{thm:esterr}. Further choose $h_{*}\asymp s^{1/2}(\log p/n_*)^{1/4}$ and $\lambda_t(t\geq 1)$ as
	\begin{equation*}
	\lambda_t\asymp \sqrt{\frac{\log p}{n_{\mathcal{A}_{\eta}}+n_0}}+\max\left\{\frac{s^2\log p}{n_*},\frac{s^3\log p}{n_{\mathcal{A}_{\eta}}+n_0}\right\}\sqrt{\frac{\log p}{n_{*}}}.
	\end{equation*}
	With the initial estimator $\tilde{\boldsymbol{w}}^{(0)}$ given by (\ref{eq:w0}) and the number of iterations $T\asymp\lceil\log((n_{\mathcal{A}_{\eta}}+n_0)/n_{*})\rceil$, the distributed QR transfer estimator $\tilde{\boldsymbol{\beta}}^{(T)}$ obtained from Algorithm \ref{alg:dist} satisfies the error bounds
	\begin{equation*}\label{eq:l1boundT}
	\inf _{\boldsymbol{B} \in \Theta(s, {\eta})} \mathbb{P}\left(\|\tilde{\boldsymbol{\beta}}^{(T)}-\boldsymbol{\beta}\|_1 \lesssim s\left(\frac{\log p}{n_{\mathcal{A}_{\eta}}+n_0}\right)^{1 / 2}+a_{\eta}+{\eta}\right) \geq 1- c_1\exp{(-c_2\log p)},
	\end{equation*}
	\begin{equation*}\label{eq:l2boundT}
	\begin{aligned}
	\inf _{\boldsymbol{B} \in \Theta(s, {\eta})} \mathbb{P}&\left(\|\tilde{\boldsymbol{\beta}}^{(T)}-\boldsymbol{\beta}\|_2 \lesssim \left(\frac{\log p}{n_0}\right)^{1 / 4}\left(\sqrt{s}\left(\frac{\log p}{n_{\mathcal{A}_{\eta}}+n_0}\right)^{1 / 4}+\sqrt{\eta}+\sqrt{a_{\eta}}\right)\right) \\&\geq 1- c_1\exp{(-c_2\log p)},
	\end{aligned}
	\end{equation*}
	where $a_{\eta}=(s\log p/n_*)^{3 / 8}\sqrt{s\eta}.$
\end{theorem}

\begin{remark}
	In comparison to the non-distributed results in Theorem \ref{thm:esterr}, a stronger condition  $\eta\ll (s\sqrt{\log p/n_0}) \wedge (n_{*}^3s\log p/n_0^4)^{1/4}$ is needed in the distributed setting for the improvement of estimation error.
\end{remark}

\bibliographystyle{biom}
\bibliography{ref}

\end{document}